\documentclass[12pt,preprint]{aastex}
\usepackage{natbib}

\newcommand{\be}{\begin{equation}}
\newcommand{\bel}[1]{\begin{equation}\label{eq:#1}}
\newcommand{\ee}{\end{equation}}
\newcommand{\bd}{\begin{displaymath}} %like \be, but doesn't put in eqn. number
\newcommand{\ed}{\end{displaymath}}   %like \ee, but doesn't put in eqn. number
\newcommand{\bea}{\begin{eqnarray}}
\newcommand{\beal}[1]{\begin{eqnarray}\label{eq:#1}}
\newcommand{\eea}{\end{eqnarray}}
\newcommand{\e}[1]{\label{eq:#1}}
\newcommand{\eqref}[1]{\ref{eq:#1}}

\def\msun{$M_\odot$}
\def\lax{{$\mathrel{\hbox{\rlap{\hbox{\lower4pt\hbox{$\sim$}}}\hbox{$<$}}}$}}
\def\gax{{$\mathrel{\hbox{\rlap{\hbox{\lower4pt\hbox{$\sim$}}}\hbox{$>$}}}$}}
\def\asec{$^{\prime\prime}$}

\begin{document}

%\slugcomment{Draft version 2; April 26, 2005;
%to be submitted to {\it The Astrophysical Journal (Letters)}.}
\shorttitle{MASS SPECTRUM OF ISM IN M82}
\shortauthors{KETO, HO, \& LO}

\title{M82, Starbursts, Star Clusters, and the formation of Globular Clusters}
\author{Eric Keto\altaffilmark{1}, 
Luis C. Ho\altaffilmark{2}, and
K.-Y. Lo\altaffilmark{3}}
\altaffiltext{1}{Harvard-Smithsonian Center for Astrophysics, 60 Garden 
Street, Cambridge, MA 02138}
\altaffiltext{2}{The Observatories of the Carnegie Institution of Washington,
813 Santa Barbara Street, Pasadena, CA 91101}
\altaffiltext{3}{National Radio Astronomy Observatory,
520 Edgemont Road, Charlottesville, VA 22903-2475}

\begin{abstract}

We observed the nearby starburst galaxy M82 in CO in the 
higher frequency (2--1) transition to achieve an angular resolution
below 1 arc second or 17 pc at the target.
We resolved the molecular gas into a large number of compact clouds, 
with masses ranging from  $\sim 2\times 10^3$ to $2\times 10^6$ \msun. 
The mass spectrum scales as $N(M) \propto M^{-1.5\pm 0.1}$, 
similar to the mass spectra of 
young massive star clusters suggesting that
individual molecular clouds are transformed in the starburst
into individual star clusters. The larger clouds are
surrounded by supernovae and HII regions suggesting that star
formation proceeds from the outside of the clouds and progresses inward
consistent with triggering by a sudden increase in external pressure.
The clouds 
with internal star formation have velocity gradients and inverse P-Cygni spectral line
profiles indicating inward motions of 35 kms$^{-1}$ consistent
with shock driven compression.
Diffuse free-free radio emission
and X-ray emission around the clouds provides evidence for
superheated ionized gas sufficient to drive the compression. 
Clouds with spectral lines indicating expansion 
show 
little internal star formation suggesting that
the dynamics precedes and is responsible for the star formation
rather than the inverse. 
M82 is known to be in interaction with neighboring M81.
The overall picture is consistent with the formation of
massive 
star clusters from individual giant molecular clouds 
crushed by a sudden galactic scale increase in external pressure
generated by the changing dynamics that result from a 
near-collision with a neighboring galaxy. Present day globular
clusters may have formed in a similar fashion in primordial 
galaxies.

\end{abstract}


\section{Introduction}

Two of the many interesting phenomena discovered by the satellite observatories
have been the starburst galaxies and the super star clusters, first 
identified as important classes by the IRAS and the 
Hubble Space Telescope (HST) satellites (Soifer et al~1987; Holtzmann et al.~1992). 
The starburst galaxies
are characterized by extremely rapid star formation
in regions with gas densities up to 20  times higher than the typical densities
of normal spiral galaxies (Sanders \& Mirabel 1996). 
Most starbursts are found in interacting galaxies, and the dynamics
of the interaction may be responsible for the high concentration
of gas and dust (Sanders \& Mirabel 1996; Kumai, Hashi \& Fujimoto 1993) 
that both provides the raw material for star formation
and also obscures and downconverts
most of the light of the starburst into the infrared.
 
In areas of lower extinction between the dust lanes of the starbursts,
recent HST observations have revealed
swarms of bright compact star clusters termed "super star clusters" or SSCs. 
The luminosity of these clusters 
comprises a significant fraction of the optical and near infrared 
light emitted in starbursts
\citep{Barth1995, Meurer1995, Maoz1996, Maoz2001, Ho1997} leading to the
inference that
the SSCs may be the dominant mode of
star formation in starburst regions (Ho 1997).
The HST observations show that SSCs represent perhaps the most extreme mode 
of star formation known. The SSCs are characterized by their 
high luminosities, 
from 1--100 times that of the R136 cluster of 30 Doradus in the
Large Magellanic Clouds, exceptional compactness (half-light 
radii $\leq 2 - 5$ pc) and young ages (\lax few hundred Myr old). 
Observations of the internal kinematics of some SSCs (\citep{HoFilippenko1996a, 
HoFilippenko1996b, Mengel2002, LarsenRichtler2004} have 
shown that these examples are extremely massive (\gax $10^5$ \msun) and
most likely gravitationally bound. Estimates of SSC masses based on population
synthesis models indicate  a
range from below $10^4$  up to a few 
$10^6$ M$_\odot$ \citep{ZhangFall1999, Melo2005}.
The similarity in size and mass of the SSCs and globular clusters have prompted 
the further suggestion that the SSCs might be present-day analogs of young globular
clusters \citep{Holtzmann1992}. 
Previously thought to form strictly in an earlier epoch 
of galaxy evolution, globular clusters
may in fact be forming in the
current epoch in starburst environments \citep{AshmanZepf2001}.

The observations suggest a fascinating connection between starbursts, SSCs,
and globular clusters, yet many questions remain. For example, 
what sets the mass scale of the globular clusters?
The mass
spectrum of globular clusters is peaked around $10^5$  M$_\odot$.  Previously this
was thought to be an observation of particular significance indicating a preferred mass 
scale at the primordial epoch of globular cluster formation (Peebles \& Dicke 1968; Fall \& Rees 1985). 
Did the globular clusters form in the past in the same way as we expect they may be
forming today, from SSCs in starbursts? Is there 
a dominant mass scale in starbursts? 
Observations of the luminosity of SSCs, translatable to
mass by means of population synthesis models, nominally show a peaked distribution similar
to the globular cluster distribution \citep{ZhangFall1999, Melo2005}.
However, the lowest luminosity clusters are not fully detectable by the HST \citep{ZhangFall1999, Whitmore1999}
and best efforts to correct the incompleteness at
the low end of the mass spectrum result in a power law spectrum rather than a peaked spectrum.
These observations suggest the hypothesis that
the ancient globular clusters formed with a scale-free power law mass spectrum, the
same as the SSCs,
and the globular clusters acquired their present characteristic mass 
after billions of years 
of evolution has destroyed the low mass end of  the original power law
\citep{Aguilar1988, ChernoffWeinberg1990, Vesperini2003}.

The mass spectrum of globular clusters at the upper end where completeness 
is  not a question has a slope of $-1.7$ \citep{HarrisPudritz1994},  
characteristic of molecular clouds in our Galaxy.
The similarity of the cloud and cluster mass spectra suggests that it may be the cloud mass spectrum
that determines the masses of the clusters.
For example, SSCs might form from the rapid transformation of entire
individual molecular clouds
into stars facilitated by the extreme environment of a starburst.
This would be as expected from simple theoretical considerations that
suggest that if a gravitationally bound molecular cloud with an equilibrium internal velocity
dispersion evolves to a stellar cluster with a similar velocity dispersion, then 
most of the cloud mass must be converted to stars
\citep{Elmegreen1983, LadaMargulisDearborn1984} 
if the cluster
is to remain gravitationally bound after the loss of whatever gas was not transformed into stars.

In addition to the possible link between starbursts, SSCs and globular clusters,
many questions hinge on the determination of the mass spectrum of the SSCs and
the molecular clouds in starbursts. 
Do SSCs form from the collapse of
individual molecular clouds (Schweizer et al.~1996; Ashman \& Zepf 2001) 
or do the SSCs form as cores in the high
densities within super-giant 
molecular clouds \citep{HarrisPudritz1994, Wilson2003}. Is this question
properly posed? For example, if the mass spectrum
of clouds is scale free, there may be no distinction between
super-giant molecular clouds and a closely packed collection of
more normal giant molecular clouds. Is star formation in starbursts
triggered  by galactic scale shocks that result from the galaxy interactions
associated with most starbursts. Is star formation sequential? 
Is the high star formation
rate in starbursts simply scaled up by the high gas density as indicated by
the universality of the Schmidt law 
(Schmidt 1959; Kennicutt 1998)? 
Is there a universal mode of 
star formation in open and bound clusters in our Galaxy and in
SSCs in starbursts such that star formation in SSCs is a scaled up version of star formation
in our own Galaxy (Elmegreen \& Efremov 1997). 
If so does this suggest a universal stellar initial mass function for all 
star formation environments.

The number of fundamental questions affected by 
the mass spectrum of molecular clouds in starbursts
motivates an observational determination.  
Molecular line observations indicate  that the molecular clouds in
starbursts are also distributed with a $-1.5$ power law spectrum.
However, 
the identification of individual clouds at the lower end of the cloud mass spectrum
is limited by the angular resolution
of the current generation of millimeter wave interferometers.
Therefore
the determination of the cloud mass spectrum,
similar to the cluster mass spectrum,   is also troubled by the completeness 
of the sample.  New molecular line observations at higher angular resolution are
required to
obtain a more complete sample at the lower end of cloud mass spectrum.

To improve on the existing molecular line
observations requires higher angular resolution. This is achievable with 
the same millimeter interferometers used in previous observations
if the new observations are made at the highest operating frequencies. 
For example, the CO(2--1) transition
at 231 GHz obtains an angular resolution (just under one arc second
at the Owens Valley Radio Observatory) that is twice as good
as that achieved by the same interferometer when observing 
the ground state transitions of CO and other molecules around 100 GHz.

We identified the starburst galaxy M82 as a suitable candidate for an
observational determination of the mass spectrum of the interstellar
medium in a starburst. First, this galaxy 
is sufficiently nearby that with arc second angular resolution
we can resolve the ISM to a linear scale
comparable to the SSC dimensions. 
M82 lies at the same distance as M81, 3.6 Mpc, which has been accurately
determined based on observations of Cepheids (Freedman et al. 1994).  At this
distance, 1\asec\ corresponds to a linear size of 17 pc, only a factor of 2
larger than the typical diameters (6--10 pc) of the SSCs. It is important to
reach the cluster scale for two reasons. First the range of the mass spectrum of clouds 
that we observe should
be comparable to that observed in clusters particularly at the critical
lower end. (The relationship between the mass and length scales is 
understood by considering that if an individual molecular cloud is
transformed by fragmentation to stars at the high efficiencies required to produce
a bound cluster, the length scales of the precursor molecular cloud and  the
resulting cluster will be comparable.) Secondly, observations at a linear
scale of 50 -- 100 pc are unable to determine whether a molecular cloud of
cluster mass has a density that is
the same as a typical molecular cloud
in our own Galaxy, or if the same mass is concentrated in a smaller cloud
of cluster scale dimensions with a density 100 to 1000 times higher, as would
be expected from
the high average surface density in the starburst environment.

Second, M82 has
a history of 
extensive SSC production, with  major 
episodes of past cluster formation and evidence for continued, ongoing 
cluster formation.  HST observations have identified
an earlier epoch of cluster formation in the region known as M82B.
(de~Grijs,
O'Connell, \& Gallagher 2001; McCrady, Gilbert, \&
Graham 2003; Melo et al. 2005; O'Connell
et al.~1995).
Current cluster formation appears to be taking place in a nearby
gas rich region where lower angular resolution observations
($> 2.5^{\prime\prime}\approx 43$ pc) 
indicate 
a large number of discrete clouds
with a power-law mass spectrum
\citep{BrouilletSchilke1993, ShenLo1995}. 
M82 is sufficiently gas-rich that it has the 
potential that
the present-day ISM might reflect, at least statistically, the initial 
conditions of the precursor clouds that might collapse to form SSCs.

While the observations were primarily motivated by the question of the mass
spectrum of the molecular clouds, a comparison of the distribution and
dynamics of the 
molecular clouds with 
the locations of 
star formation activity as indicated by other observations, suggests 
several interesting hypotheses about star bursts and the formation
of massive star clusters. The observations suggest that
star formation begins in the outer 
regions of the clouds rather than their centers. This is different than star formation in
our own Galaxy where we find 
star formation associated primarily with the centers of molecular clouds
where, because of the self-gravity of the clouds, the gas density is highest.
A comparison of star formation activity and the gas velocities in the clouds in M82, 
as revealed by the spectral information in our observations,
shows that star formation occurs where the velocity fields indicate massive compression
on the scale of the giant molecular clouds.
Theoretical considerations then lead to a hypothesis for star formation in star bursts
triggered by large scale shocks driven into the molecular clouds 
by a sudden increase in the external pressure. 
This mode of star formation is fundamentally different from 
normal star formation in our own Galaxy that takes place more slowly 
presumably motivated by a gradual increase in the density and pressure of the ISM
or the slow loss of the internal energy of molecular clouds.

\section{Observations and Data Analysis}

The observations were made using the Millimeter Array at the Owens Valley Radio 
Observatory (OVRO). Approximately 64 hours of data were collected  on a number 
of dates during the winter seasons of 1999 and 2000.  Data from two phase 
centers, the left and right sides of the galaxy located at ($\alpha, 
\delta$)$_{\rm J2000}$ = ($09^h 55^m 53.19^s$, $69^\circ 40' 51.9''$) and
($09^h 55^m 48.33^s$, $69^\circ 40' 44.0''$), respectively, were combined to 
form an image of the central $ 1 \times 0.5$ kpc of M82  made up of 
2 overlapping beams
of 25\asec\ diameter separated by 25\asec .  
The correlator was tuned to
230.53799 GHz, the rest frequency of CO(2--1), with a spectral resolution of
4 MHz ($\sim$5 km s$^{-1}$) in 128 frequencies. The channels were Hanning
smoothed in the data processing to achieve a final spectral resolution of
10.4 kms$^{-1}$ per channel. Baselines ranged from 20 to 
180 k$\lambda$ covering the range of spatial scales from approximately
1$^{\prime\prime}$ to 10$^{\prime\prime}$ or 17 to 170 pc in linear
scale at the distance of M82.  Calibration, done in the MIR data reduction package 
developed at OVRO and the Submillimeter Array, used the Seyfert galaxies 
NGC 1275 (3C 84) and 0923+392 (4C 39.25) for amplitude and gain calibration,
and the quasar 3C 273 for bandpass calibration.  Mapping was done with the 
NRAO AIPS software.
The FWHM of the synthesized beam was $1.0^{\prime\prime} \times 
0.8^{\prime\prime}$, and the rms noise per channel was 0.01 Jy beam$^{-1}$.  

The integrated intensity of the CO(2--1) emission above a threshold of
twice the rms noise or $2 \sigma = 0.02$ Jy beam$^{-1}$
is shown in figure \ref{fig:integrated_intensity}.   
The map shows a number of large cloud complexes along the major axis of the 
galaxy, which is viewed nearly edge-on.  The smallest clumps depicted in the 
map are unresolved.  

The structures seen in our high resolution observations of CO(2--1) are consistent 
with the structures seen in previous molecular line observations, but there
are differences.
Because our observations were made 
at high angular resolution with an interferometer sensitive to a limited
range of spatial scales, they do not have the same sensitivity to large scale 
diffuse structure as previous observations made at lower angular resolution.
Secondly,  the spatial distribution of CO(2--1) more closely matches 
HCN(1--0) (Brouillet \& Schilke 1993), than
CO(1-0)
(Shen \& Lo 1995; Matsushita et al.~2005).
Because HCN  has
a higher critical density for excitation than CO(1--0),
the comparison suggests that the CO(2--1) traces denser gas than CO(1--0).
The combination of sensitivity to denser gas and the lack of sensitivity
to diffuse extended structure is evident in a comparison of our map of the
CO(2--1) integrated intensity with
the map of CO(1--0) integrated intensity
(Matsushita et al.~2005).
The structure in the CO(1--0) integrated emission
that defines the molecular wall of a superbubble in the nucleus of
M82 (Matsushita et al.~2005) is not identifiable as such in our map of CO(2--1) integrated emission.
The high velocities of the molecular gas defining the boundary of the superbubble are seen
in our data, and these velocities are consistent with molecular gas in an expanding  bubble
(Matsushita et al.~2005)
rather than molecular gas in a spiral arm (Shen \& Lo 1995).
There is no report in the literature of the total emission in CO(2-1) that
could be used to quantify how much
of the large scale structure  our interferometric observations are missing.

\section{The Mass Spectrum of Clouds and Clusters}

\subsection{Details of the Observational Determination of the Mass Spectrum of the Clouds}

The CO observations are of course a projection of the emission on the plane
of the sky and a determination of the three dimensional distribution of the molecular
gas so that individual clouds can be identified is an interesting research
topic itself (Williams, de Gues, \& Blitz 1994; Bensch, Stutzki, Ossenkopf 2001; Stutzki \& Gusten 1990).
Here we rely on previous research and use the
well defined algorithm {\tt clumpfind} as detailed in Williams, de Gues, \& Blitz (1994).
This algorithm uses contours of the CO emission to define the boundaries of
individual clouds on the two dimensional map plane, and the gas velocities
as determined by the spectral line information to define the boundaries on
the third dimension.
We used {\tt clumpfind} to identify individual clouds to a 
threshold of twice the rms noise in a single channel as suggested by Williams, 
de Gues, \& Blitz (1994). Because the map of integrated intensity (figure  1) was 
also made with a threshold of 
$2\, \sigma$,
all the clouds seen in figure 1 represent at least one clump as determined by
the {\tt clumpfind} algorithm. A total of 303 discrete clouds were found by the 
algorithm.  

Once the boundaries of the clouds have been identified, the determination
of the enclosed mass, mostly H$_2$, from the information provided by the CO tracer
is another interesting research topic.
Recent estimates of the conversion factor between 
the integrated emission of CO(1--0) and the column density of H$_2$ 
in galaxies vary by a 
factor of 10, and may depend on whether the galaxy is a starburst. 
\cite{Rosolowsky2003} estimate $2\times 10^{20}$ H$_2$ cm$^{-2}$ 
(K kms$^{-1}$)$^{-1}$ for M33, a relatively normal spiral. This conversion 
factor is the same as that estimated for our Galaxy \citep{Strong1988}. For 
starburst galaxies, \cite{Davies2004} estimate a factor of 0.4 to 0.8 times
the Galactic value, while other studies find
factors ranging from 0.2 to 0.1 
\citep{Wang2004, Wada2005}. 
Matsushita et al.~(2000) suggest a conversion factor of $1.4 \pm 0.6 \times 10^{20}$.
We adopt
a value of $1.0\times 10^{20}$ H$_2$ cm$^{-2}$ (K kms$^{-1}$)$^{-1}$.

The empirical conversion factors above are all determined for the integrated intensity of CO(1--0) 
whereas our observations are of CO(2--1).  Because the conversion factor is 
essentially empirical, there is little theoretical guidance as to how one should modify this 
factor for the (2--1) line.  Depending on the excitation temperature 
the population in the (2--1) line may be different from that in the (1--0) line.
Excitation temperatures can be 
estimated from observed ratios of certain molecular lines.   
Temperatures derived for the clouds in M82 
are $11.5 K \pm 3.0$ from cyclopropenylidene in a 30\asec\ beam 
(Oike et al. 2004), 30 K from NH$_3$ in a 40\asec\ beam (Mauersberger et al. 
2004), and $50 \pm 20$ K from CO in a 20\asec\ beam (Petitpas \& Wilson 2000).
We can bracket the problem. For excitation temperatures between 5 
and 50 K, the line ratios in local thermodynamic equilibrium vary between 0.5 
and 1.5. Thus over this range of temperature, a population ratio of unity should 
be within a factor of 2 and in agreement with  a line ratio of 0.9 found empirically in a CO survey of nearby
galaxies (Braine \& Combes 1992).
The difference in brightness of the two lines owing 
to the frequency dependence of the source function will be accounted for in 
the conversion from Jy~beam$^{-1}$ to K, measured to be 23 Jy K$^{-1}$ in our calibration.

While a determination of the absolute cloud masses from the CO line emission is 
known to be uncertain \citep{Wilson1995, Israel1997},  the direct 
conversion still provides an indication of the masses. 
Furthermore, the 
relative masses, and therefore the slope of the mass spectrum, should be less affected by 
conversion uncertainties.  
The formal statistical uncertainty in the mass 
as derived from the propagation of the error in the measured flux in each 
pixel is relatively small compared to the uncertainty in the conversion 
factor. The average of the formal error of each of the clouds is  2\%, while
the uncertainty in the conversion factor is on the order of at least a factor 
of a few.

The mass derived from the integrated intensity of CO may be compared to the
virial mass derived from the length scale and the internal velocity dispersion.
The use of the virial theorem in estimating the masses of molecular clouds has
a long and successful history in studies of the ISM.  However, recent
theoretical work suggests that interstellar clouds may not be gravitationally
bound entities in equilibrium, but rather transient structures in  turbulent
flows (Larson 1981; MacLow \& Klessen 2004; Elmegreen \& Scalo 2004). 
If so, the concept of virial equilibrium may not apply to interstellar
clouds. However if the turbulence itself is driven by gravitational forces,
that is if the velocities are primarily driven by the
gravitational potential of the mass of the ISM, then there would 
be approximate equipartition of kinetic and potential 
energies within the flow \cite{Larson1981}. In this case, even though the
clouds are not in virial equilibrium, 
the virial relationships should still apply. 
We use the equation for the virial mass,  equation A3 
in \cite{Williams1994}, that expresses the mass in 
quantities reported by the {\tt clumpfind} algorithm,
$M_{\rm vir} = 5\Delta R \sigma_v^2/\alpha G$, where $\Delta R$ is the cloud 
radius, $\sigma_v$ is the line-of-sight internal velocity dispersion, and 
$\alpha$ is a geometrical factor equal to 5/3 for a density profile scaling as 
$r^{-2}$.  
The radius derived from the {\tt clumpfind} algorithm has been corrected for  the
spatial resolution following Equation A7 of \cite{Williams1994}.  Clouds 
that have clump radii less than the beam size have been set to a size 
equivalent to a single pixel, or $0.2\times 0.2$ square arcseconds. 
The data are not similarly corrected for the spectral resolution because the
channel width of the data is the same width and function as the spectral resolution. 
Because of the finite spatial and spectral resolution, the masses of some 
of these clouds may be upper limits. The formal error on the virial mass, based 
on an assumed error in $\Delta R$ equal to the half-width at half maximum 
of the beam and an assumed error in $\sigma_v$ of half the channel width,
is 100\%. The error in the virial mass is thus of the order of the uncertainty in 
the conversion mass.

Figure \ref{fig:virial_mass} shows that with the adopted CO-to-H$_2$ mass conversion, 
the clouds in M82 follow the relationship of virial equilibrium.
The lowest-mass clouds, below $10^4$ $M_\odot$, appear on average to be 
slightly more massive than required for equilibrium.  If one excludes the 
clouds with masses below $10^4$ $M_\odot$, the fit is again indistiguishable 
from virial equilibrium.  This result confirms, a posteriori, the  adopted
CO-to-H$_2$ conversion factor and suggests that the conversion factor
in M82 differs from 
the standard Galactic value by a factor of 2.

A histogram of cloud masses is shown in figure \ref{fig:mass_spectrum}, binned on a
scale of $\Delta \log M = 0.165$ that provides 20 bins of equal logarithmic size
over the logarithmic mass range of the clouds from 3.4  to 6.7  log M$_\odot$. 
Over this mass range, the mass spectrum 
can be reasonably described by a single power law, d$N$/dlog$M \propto 
-0.5\pm 0.04$, that on a linear scale corresponds to $N(M) \propto 
M^{-1.5\pm 0.04}$.
The least squares fit to the slope includes all the clouds 
and assumes that the data $N(\log M)$ have equal weight.  

The histogram includes all the clouds found by the {\tt clumpfind} algorithm.
Since the interferometer is only sensitive to emission on 
scales of 1$^{\prime\prime}$ to 10$^{\prime\prime}$, the mass spectrum 
at either end of the range of spatial scales 17 to 170 pc is uncertain.
Thus the completeness of the sample may be primarily dependent on the sensitivity to angular
scale rather than, as in the optical case \citep{Whitmore1999},
on the confusion with the background emission. 
In interferometric observations
the larger scale
background emission  is essentially invisible

At the lower end of the mass range, where the cloud 
sizes are below the resolution limit of the interferometer, the 
finite resolution of the interferometer causes 
both an increase in the apparent size of the cloud as well as an approximately 
offsetting decrease in measured antenna temperature.  Thus these two errors
may cancel resulting in higher uncertainty rather than a trend.
Nonetheless, if one wishes to 
disregard the clouds at the lower end of the mass range, setting 
the lower mass limit to $10^4$ $M_\odot$ results in a slope of $-1.6 \pm 0.06$ 
rather than $-1.5 \pm 0.04$. At the upper end of the mass range, the expected
errors are somewhat different.
The masses of the clouds at the upper end of the mass range may represent 
lower limits to true masses if the sizes of the clouds exceed the angular scale of 
$\sim$10\asec\ that is captured by the interferometer. There is no way of 
estimating the mass or structure of the clouds beyond the angular limit from 
the data set. However, this uncertainty affects only the most massive clouds. 
A fit to the histogram ignoring clouds with masses $>10^6$ \msun\ results
in a slope of $-1.4 \pm 0.05$.
The different slopes of the  mass spectrum derived with the different mass limits 
suggests that the errors are dominated by uncertainties at either end of the 
mass range.  Thus, the slope of the mass spectrum may be approximately
estimated as $-1.5 \pm 0.1$.

\subsection{Comparison of the Mass Spectrum of the Clouds and Clusters}

The mass spectrum of the molecular clouds in M82 
over the mass range that we observe with the interferometer, $\sim 10^3 - 10^7$ M$_\odot$, 
is well approximated
by a power law with a slope of $-1.5$. 
The nominal mass spectrum of the SSCs in M82, calculated from continuum magnitudes measured
by the HST, and
taking into account an estimated extinction and age, but
uncorrected for completeness, 
is peaked at a mass of about $10^5$ M$_\odot$
\citep{Melo2005, deGrijs2001}.
Investigation of the completeness of HST observations of SSCs in the Antennae galaxies suggests that
the peak in the distribution is due to the incomplete detection of faint clusters
below the peak \citep{Whitmore1999}. Correction for the incompleteness results
in a power law distribution for the luminosity.
If the peak in the mass spectrum of the clusters in M82 is due to incompleteness,
we may still derive the slope of the power law distribution by fitting the
upper end of the cluster mass spectrum where the data are assumed to be complete. 
We derive a slope of $-1.4$ (figure \ref{fig:melo_mass} in our paper) from the masses of all the clusters
listed in \cite{Melo2005} (tables 5 -- 9). (This power law distribution in the upper
end of the mass range of the M82 clusters
is similar to the power law mass spectra measured
in other galaxies such as the Antennae and M51 (Zhang \& Fall 1999; Bik et al.~2003)).
Thus the slopes of the mass spectra of the clouds and clusters in M82 are consistent with
each other and consistent with the hypothesis that the mass scale of the molecular clouds determines
the mass scale of the clusters.
Furthermore,
the slope of the mass spectrum of 
clouds in M82 is comparable to the slopes of the upper end of the mass 
spectra of old globular clusters in a  number of galaxies \citep{HarrisPudritz1994}
including M87 ($\alpha=-1.60$), NGC1399 ($\alpha=-1.61$), NGC4594 ($\alpha=-1.54$), 
and NGC4636 ($\alpha=-1.90$).
If the lower end of the mass spectrum
of the globular clusters can be assumed to be determined by the destruction
of low mass clusters, then the mass spectra of the clouds, SSCs and globular
clusters are consistent with the hypothesis of
a common origin for globular clusters and SSCs.

\section{Comparison with Indicators of Star Formation}

The relationship between the dense molecular clouds
and star formation may be explored by comparing
the CO emission with several
indicators of star formation such as the SSCs themselves,
radio point sources, diffuse radio free-free emission, and mid-IR emission. 
Each of these observables provides different 
information on the star formation activity because of their 
relationships to different emission mechanisms and
because of the different extinction at the different
wavelengths.

The SSCs, as identified by HST I-band and H$\alpha$ emission 
\citep{Melo2005},
are seen to be spread throughout the nuclear region except at the locations of
the brightest CO emission in the larger clouds (figure \ref{fig:ssc_co}). 
Because dense molecular gas 
obscure the optical light from the SSCs,
the inverse correlation between the SSCs 
and the CO emission does not
necessarily imply a lack of clusters in the molecular clouds. 
But the presence of the clusters around the clouds certainly indicates
recently completed star formation around the clouds. We know that the
star formation is essentially finished in these optically visible clusters 
because if there were dense gas within the clusters available for 
further star formation, 
then this gas within the clusters would also obscure much of the optical emission from the clusters.

The centimeter-wavelength radio point 
sources \citep{Rodriguez-Rico2004} are associated with
supernova remnants and 
H~{\sc ii} regions and the two may be distinguished by their radio spectral indices. In
addition the HII regions may have
detectable recombination line emission. 
Although the number of radio point sources and the number of bright
clouds are both few, the radio point sources in figure \ref{fig:comparison} appear
clustered around the edges of
the clouds, but  absent in 
the centers of the brightest CO clouds
with the exception of the one large cloud at
9 h 55 m 51.4 s +69 d 40 ' 44 " (marked A on figure \ref{fig:integrated_intensity}) that is coincident with
several HII regions.
If this inverse correlation
is significant, then
because the centimeter emission is little affected by extinction, the absence
of supernovae and HII regions within the molecular clouds indicates a lower
rate of star formation in the centers of most of the clouds than on their periphery.

If this peripheral star formation is currently ongoing, it should be occurring in
molecular gas. The lack of detection in our CO observations indicates that this
gas mst have a characteristic scale exceeding 10" that is beyond the range of spatial
scales detectable by the interferometer. Alternatively, the star formation may have
consumed and dispersed the molecular gas around the dense clouds. In this case, the
star formation must have just finished.

Mid-infrared emission indicates the presence of warm (T \gax 100 K) dust
and is often associated with massive star formation. 
A comparison of the mid-IR emission (Lipscy \& Plavchan 2004) with the
CO emission shows an association, but again a lack of detailed correspondence (figure \ref{fig:midIRvelocity})
with the exception of the large cloud at 9 h 55 m 51.4 s +69 d 40 ' 44 "
(marked A on figure \ref{fig:integrated_intensity}).
and a small cloud to the west (right) at  9 h 55 m 51.4 s +69 d 40 ' 44 " (marked B on figure \ref{fig:integrated_intensity}).
The mid-IR emission within the large cloud "A" is not coincident with the
peaks of the CO emission but forms an arc just to the south and west
around the brightest CO peak.
Since the mid-IR emission is little affected by extinction,
the weak mid-IR emission in the centers of most of the clouds 
again suggests less  star formation activity in the
centers of the most of the CO clouds. 

Matsushita et al.~(2005) suggest that the diffuse 100 GHz continuum emission in M82
is a tracer of star formation activity
because it is dominated by free-free emission from gas that has been ionized
by supernovae-driven shocks and the radiation from massive stars.
The comparison (figure \ref{fig:comparison}) between the continuum and the CO emission
indicates star formation in the same two regions as indicated by the mid-IR although
the 100 GHz continuum observations do not have the angular resolution to 
determine the location of the star formation within the largest cloud.
Similar to the mid-IR, the 100 GHz continuum indicates
star formation around all the molecular clouds, but no increase at the
locations of most of the clouds.
Because the centimeter radio emission is little
affected by extinction, the lack of continuum emission in the centers of most
of the clouds indicates no increase in star formation within these clouds.

In summary, 
the comparisons of the CO with several tracers of star formation activity 
fail to find evidence for a correspondence between current star formation and the
column density of CO. Only one of the large clouds shows star formation activity
in its interior and then only on one side, just off the center. Only one smaller cloud
shows evidence of interior star formation aligned with the CO peak.
Instead current star formation  generally appears around the edges of most of the largest clouds.
The several comparisons can be rationalized into a hypothesis of inwardly progressing
star formation.
The distribution of
SSCs around the dense molecular clouds suggests that star formation has occurred
in the surrounding area over the past 25 Myr with about half the clusters formed
in the last 6 Myr \citep{Melo2005}.  
The distribution of supernovae and
HII regions suggests that star formation is occurring on the periphery of the
giant molecular clouds at the current time.
If the centers of molecular clouds are sites of future star formation, the observations
are consistent with star formation propagating inward through the molecular clouds
from the edge toward the center.
This inwardly propagating star formation is on the scale of individual clouds (\lax 50 pc)
and is a smaller scale than that of the outwardly propagating star formation (500 pc)
suggested by Satyapal et al.~(1997). The two hypotheses are quite consistent with
each other if an earlier generation of star formation triggers further star
formation by compression of the molecular clouds.

\section{Triggering the star formation}

\begin{figure}[t]
\vskip -1.0truein
\includegraphics[width=5truein,angle=90]{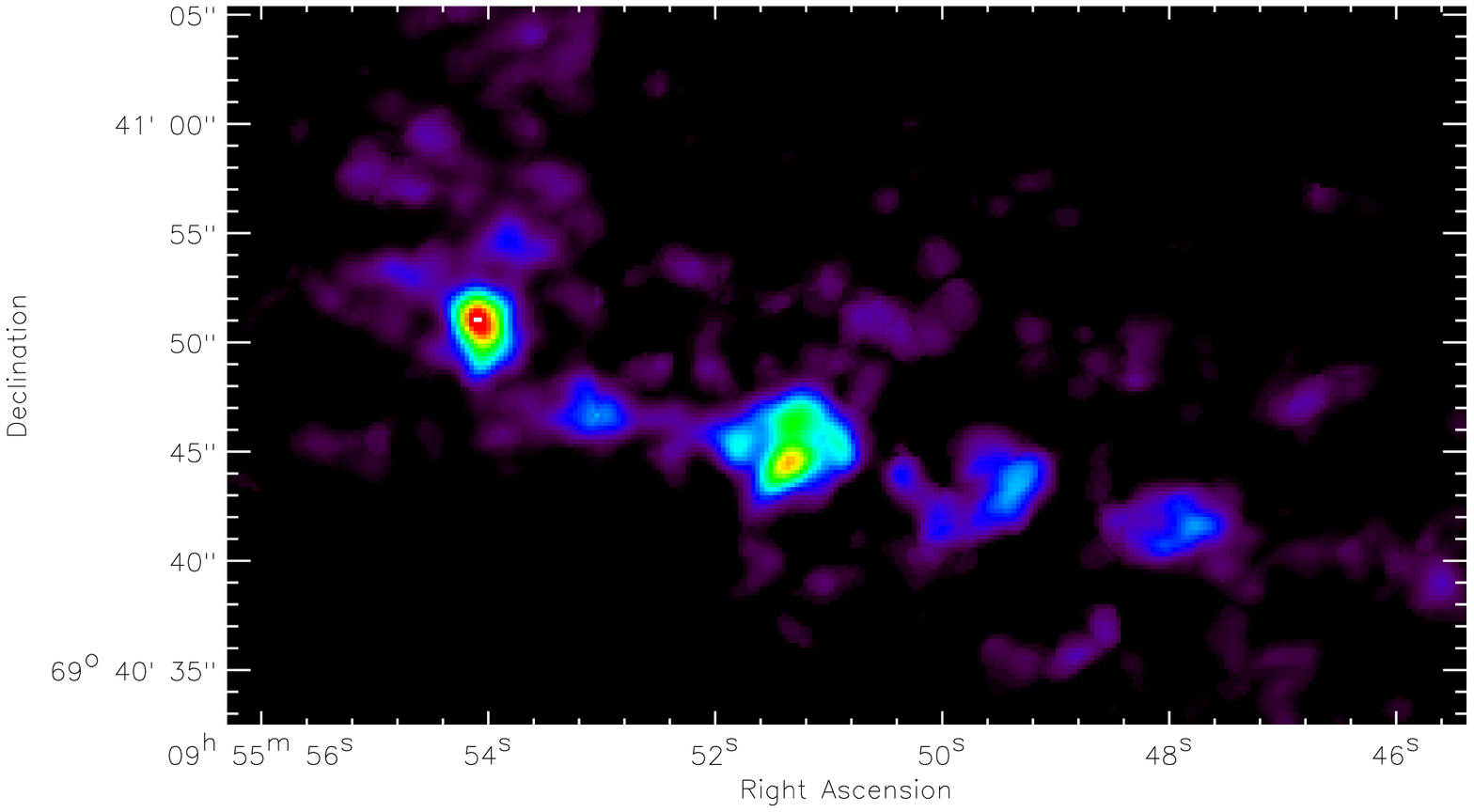}
\vskip -2.0truein
\includegraphics[width=5truein,angle=90]{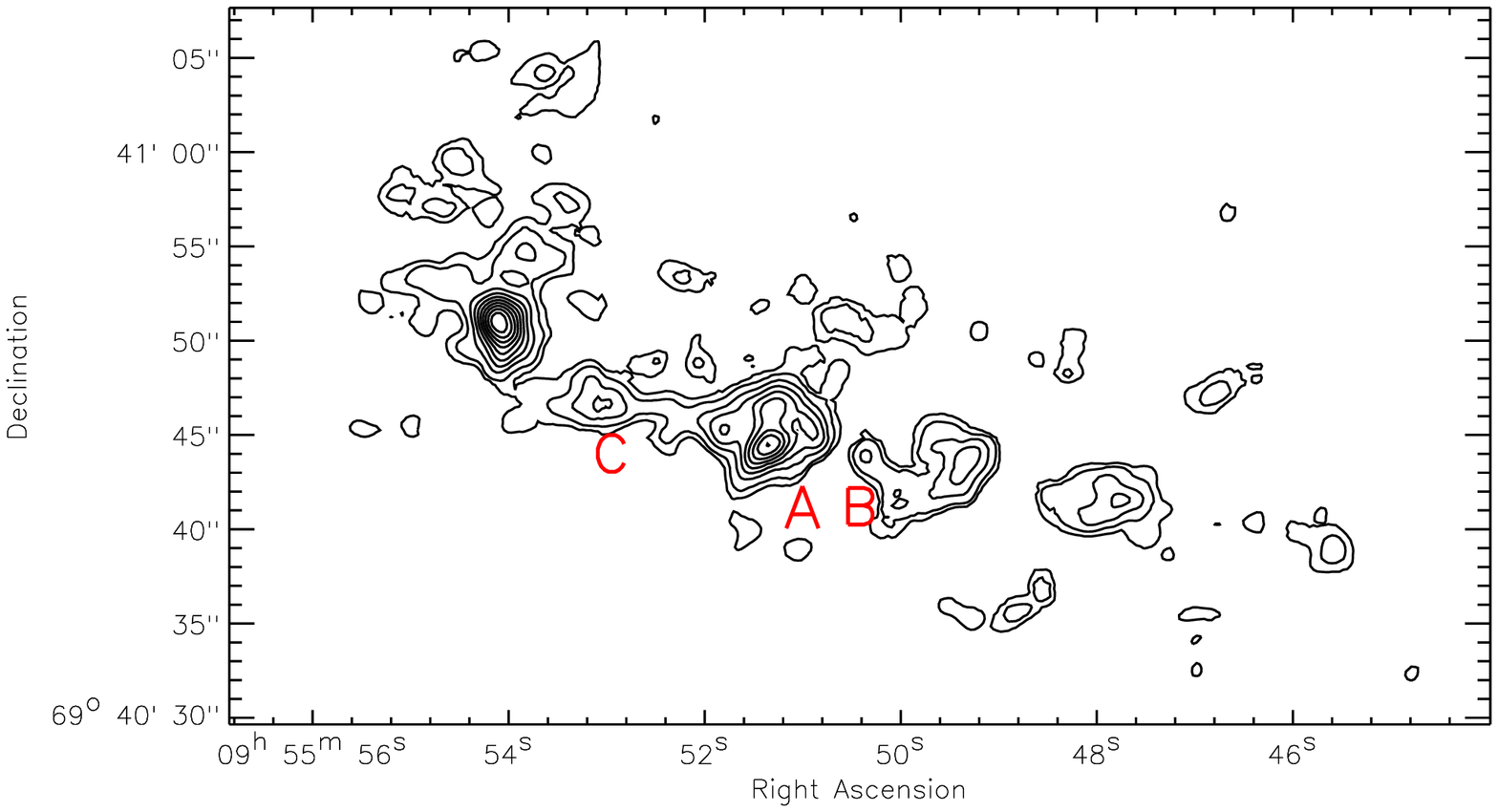}
\vskip -1.0truein
\caption{
The integrated intensity of CO(2--1) emission in the central region of M82
in color and in contour.
The contours are in multiples of 10\% of the peak flux of $8.57 \times 10^4$ 
Jy beam$^{-1}$ km s$^{-1}$; an extra contour is drawn at the 5\% level.  
The epoch of the coordinates is J2000.
}
\label{fig:integrated_intensity}
\end{figure}

\begin{figure}[t]
\includegraphics[width=5truein]{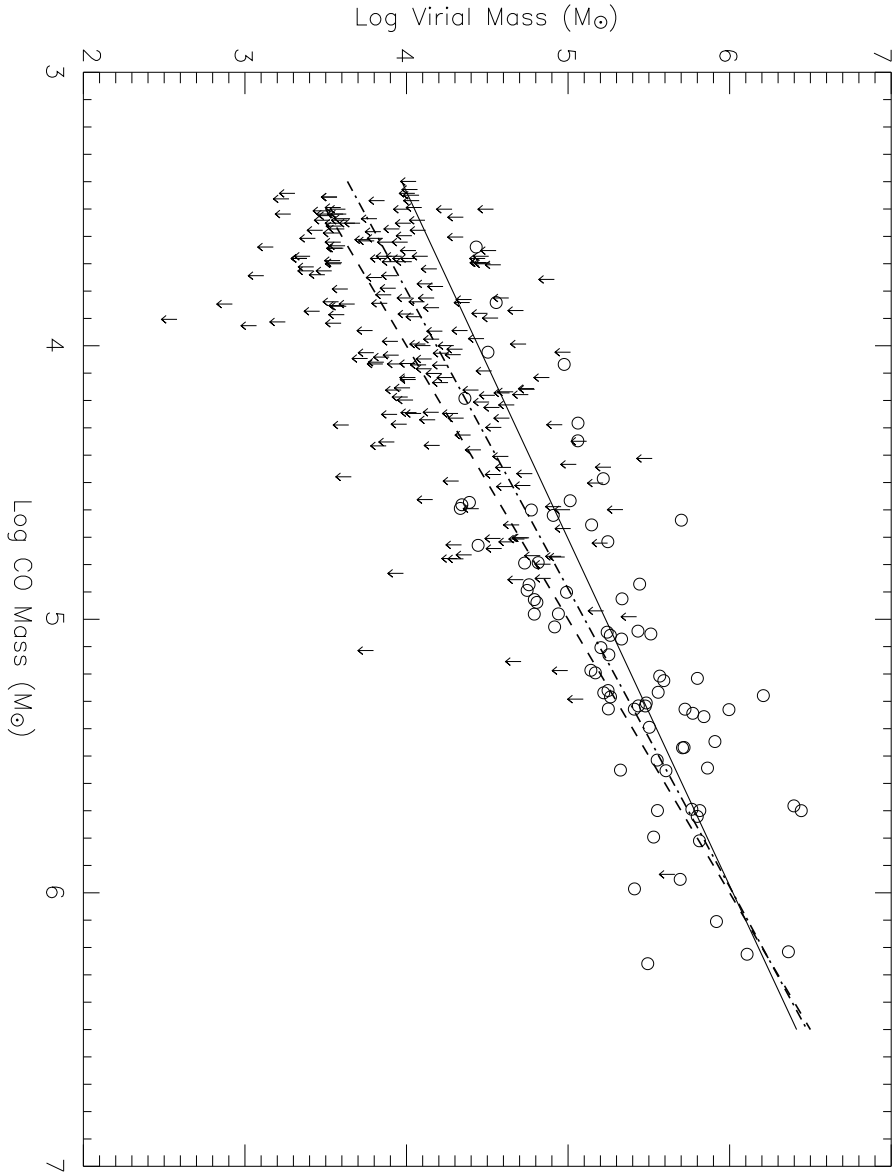}
\caption{
The virial mass ($M_{\rm vir} = 5\Delta R \sigma_v^2/\alpha G$) compared with
the mass determined from the CO line brightness.  Clouds with unresolved
sizes are plotted as upper limits.  If the error on the mass of each cloud
were 100\%, the error bars would have a width of 0.3 in each direction.
The {\it dashed line}\ is the line of virial equilibrium (virial mass =
CO mass). The best fit through the masses of the resolved clouds, those
with radii larger
than the angular resolution, is shown as a {\it solid line} while the
best fit through all the clouds is a {\it dot-dashed line}.
}
\label{fig:virial_mass}
\end{figure}

\begin{figure}[t]
\includegraphics[width=5truein]{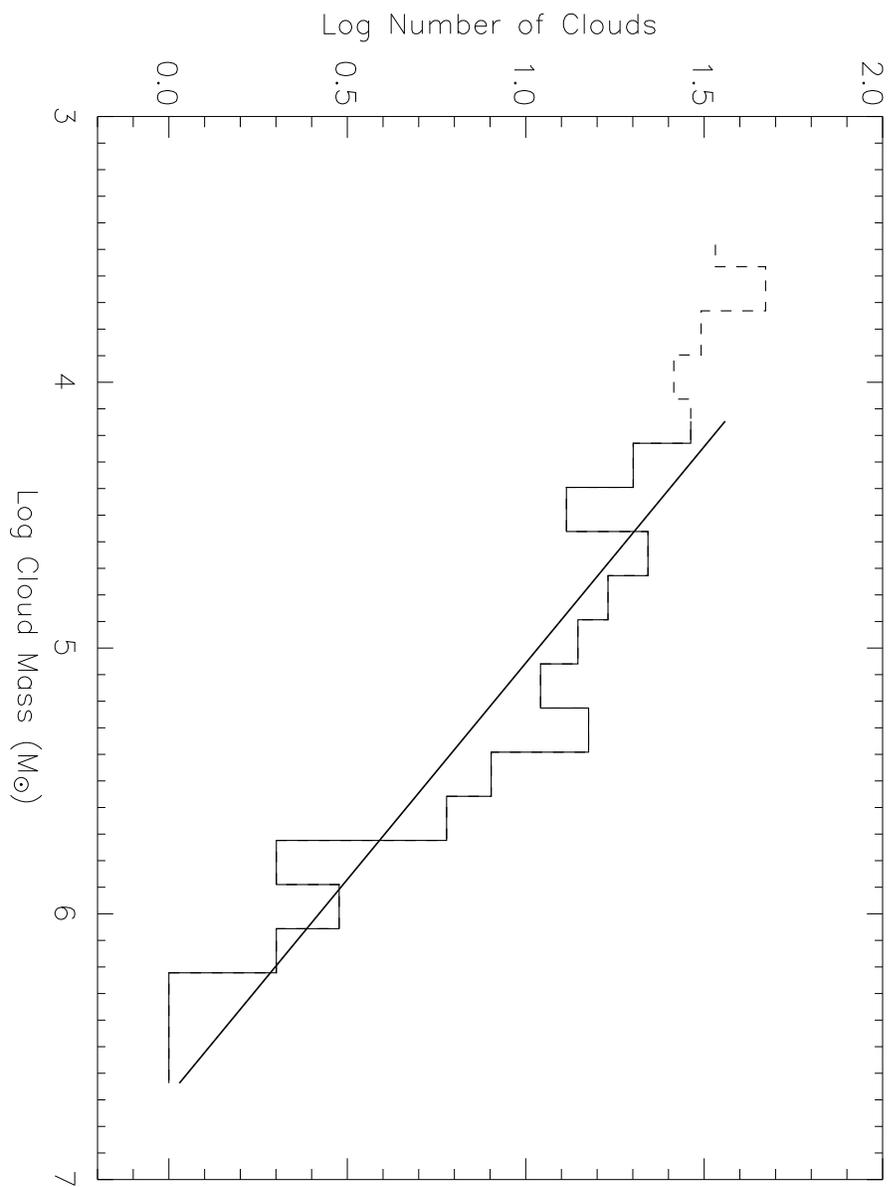}
\caption{
The mass spectrum of molecular clouds in the nucleus of M82.  The cloud masses
were estimated from the CO brightness, and are binned at
$\Delta \log M = 0.165$ to produce 20 bins across the mass range from
3.4 to 6.7 log M$_\odot$.  
The heavy solid line is a fit to the M82 data,
The data from the dashed portion of the histogram, cloud masses $< 10^4$ M$_\odot$,
was not included in the fit.
$dN/d\log M \propto -0.5\pm 0.04$.
}
\label{fig:mass_spectrum}
\end{figure}

\begin{figure}[t]
\includegraphics[width=5truein]{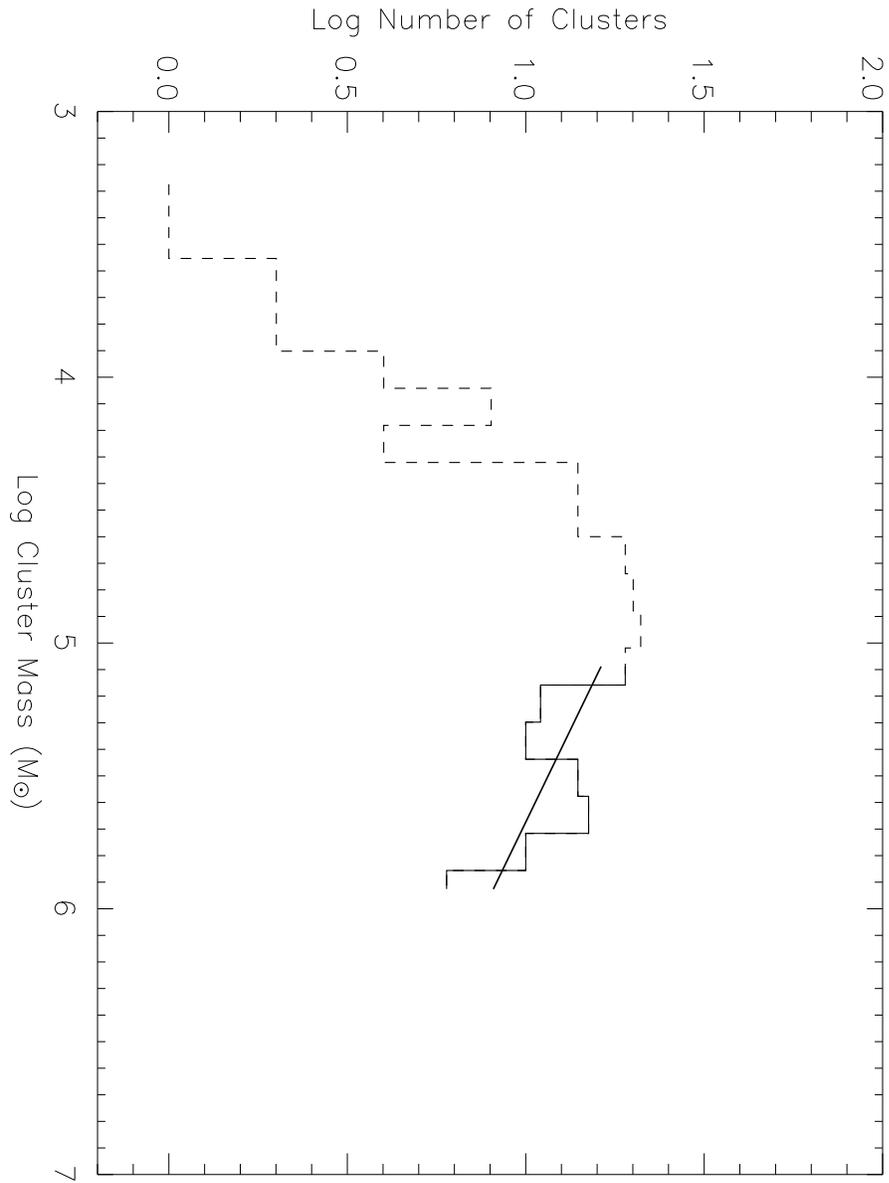}
\caption{
The mass spectrum of the SSCs in the nucleus of M82
from \cite{Melo2005}.
The histogram shows the mass spectrum. The slope of 0.4 (d$N /dlog$M) is fit to the
upper end of the mass spectrum where the sample is assumed to
be complete. The data on the dashed portion of the histogram
were not used in the fitting.
}
\label{fig:melo_mass}
\end{figure}

\begin{figure}[t]
\includegraphics[width=5truein,angle=90]{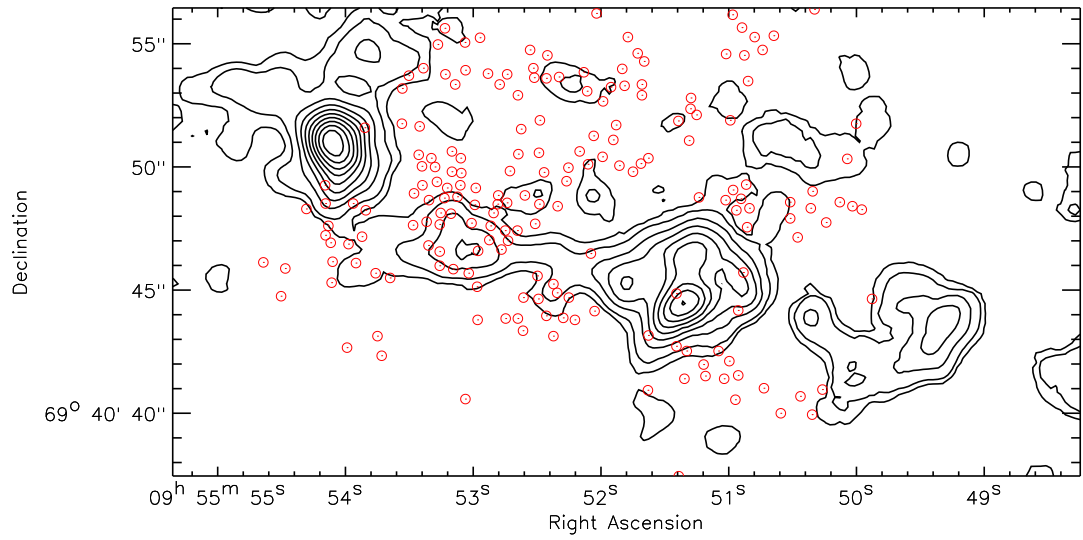}
\caption{
The CO(2-1) integrated intensity with young SSCs
from \cite{Melo2005}. The observations of the SSCs covered only a
portion of the center of the galaxy, approximately as outlined
by the extent of the SSCs.
The epoch of the coordinates is J2000.}
\label{fig:ssc_co}
\end{figure}

\begin{figure}[t]
\vskip -1.0truein
\includegraphics[width=4.5truein,angle=90]{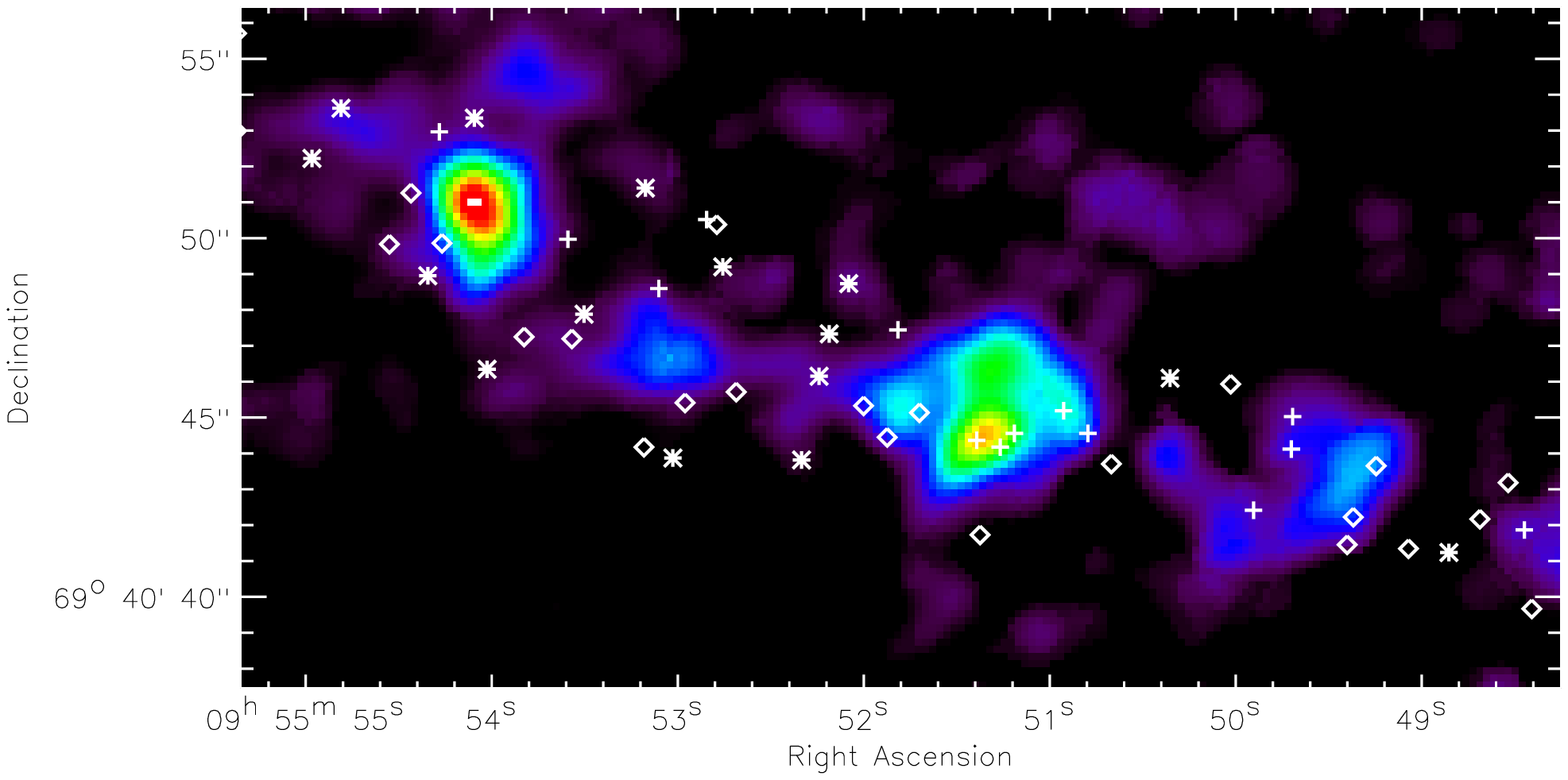}
\vskip -2.0truein
\includegraphics[width=4.5truein,angle=90]{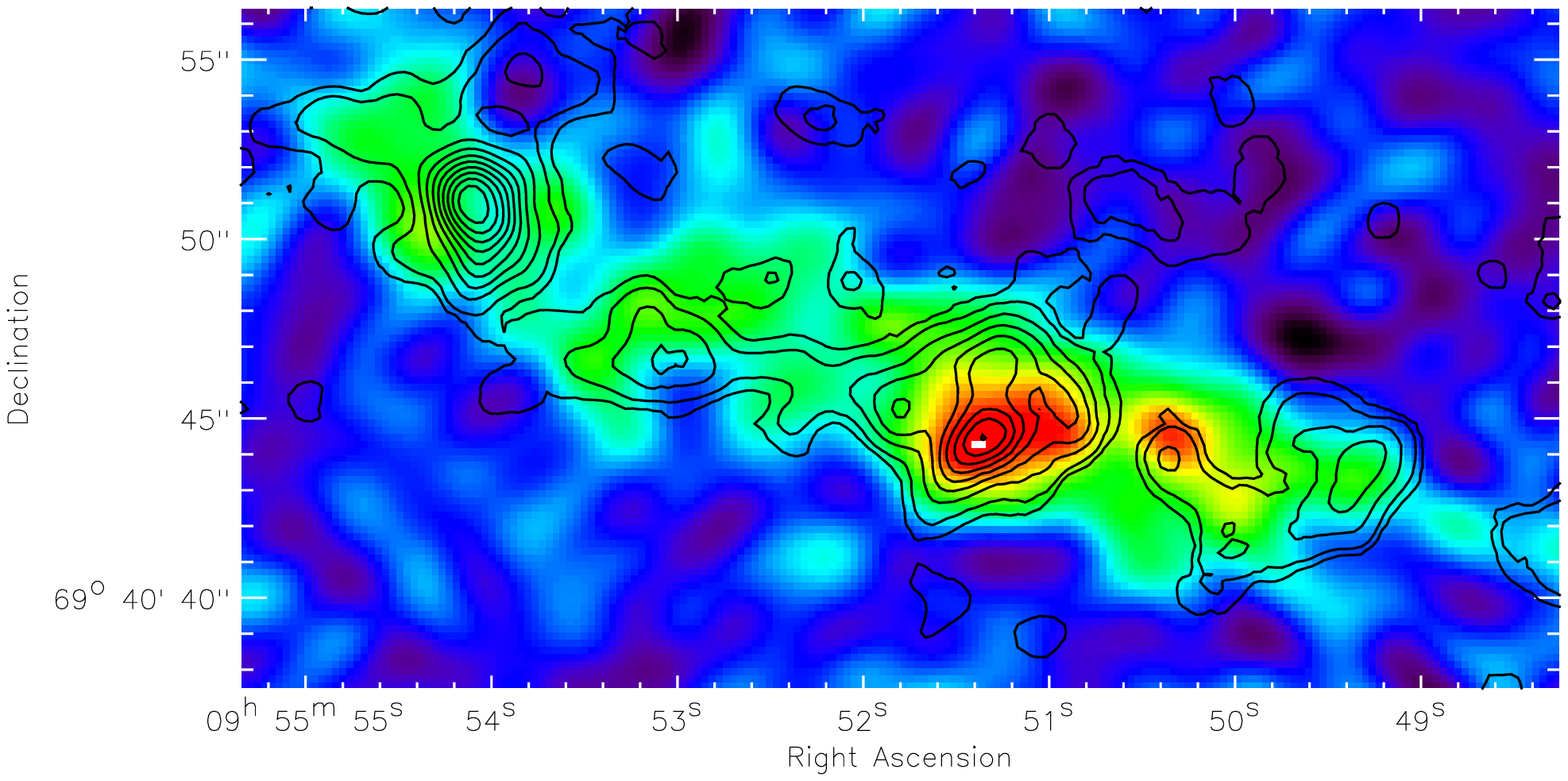}
\vskip -1.0truein
\caption{
The integrated intensity of CO(2--1) emission in the central region of M82.
The symbols show the 
positions of radio point sources from \cite{Rodriguez-Rico2004}.  
Diamonds represent supernova remnants, 
crosses represent H~{\sc ii} regions, asterisks represent radio point sources 
of unknown spectral index (either supernova remnants or H~{\sc ii} regions).
\hfill\break
The integrated intensity of CO(2-1) in contour and the 100 GHz radio continuum
from \cite{Matsushita2005} in color. 
The
CO(2-1) is generally inversely correlated with 
the radio point sources not correlated in detail with the extended radio continuum.
The epoch of the coordinates is J2000.
}
\label{fig:comparison}
\end{figure}

\begin{figure}[t]
\vskip -1.0truein
\includegraphics[width=4.5truein,angle=90]{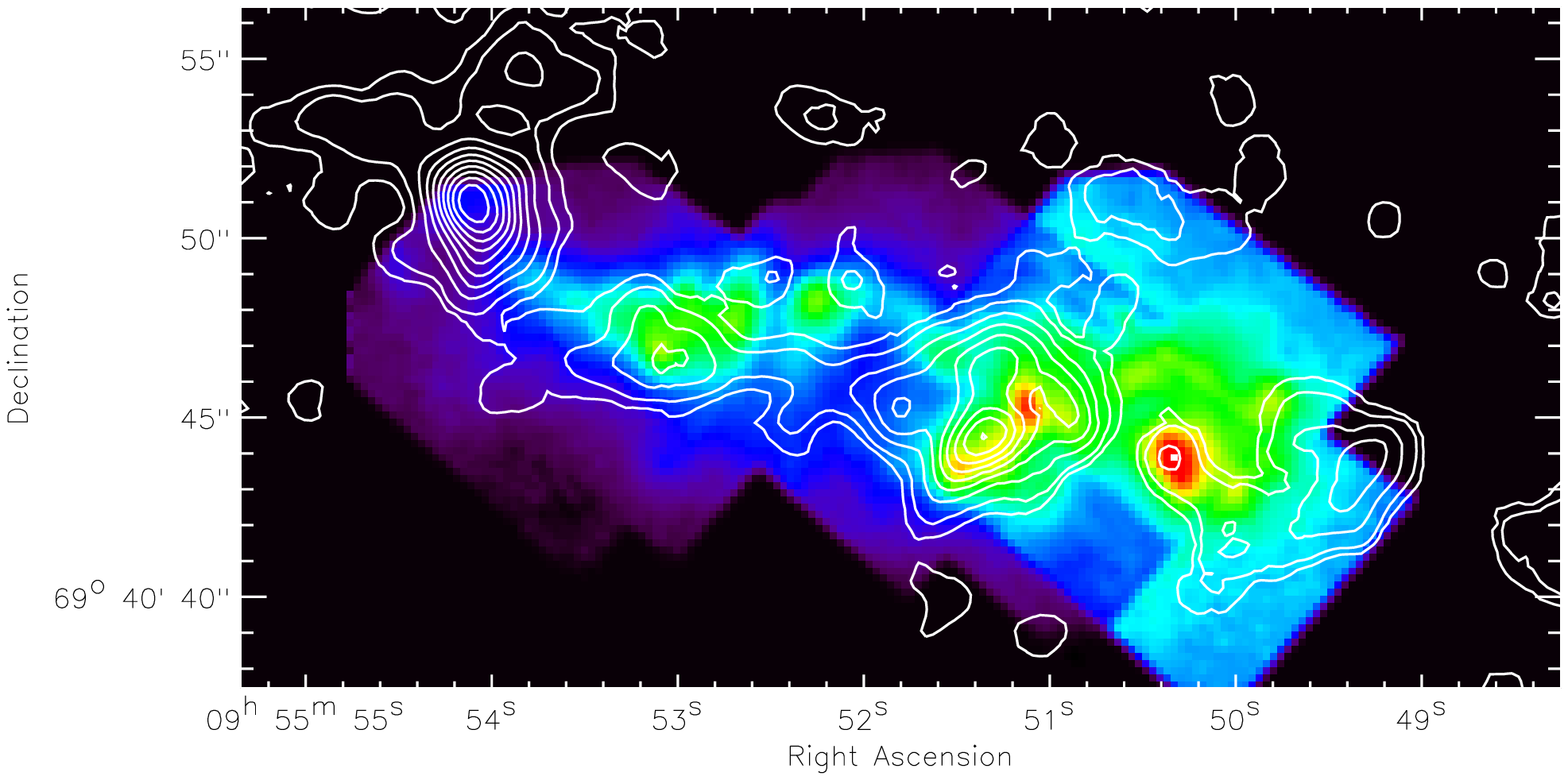}
\vskip -2.0truein
\includegraphics[width=4.5truein,angle=90]{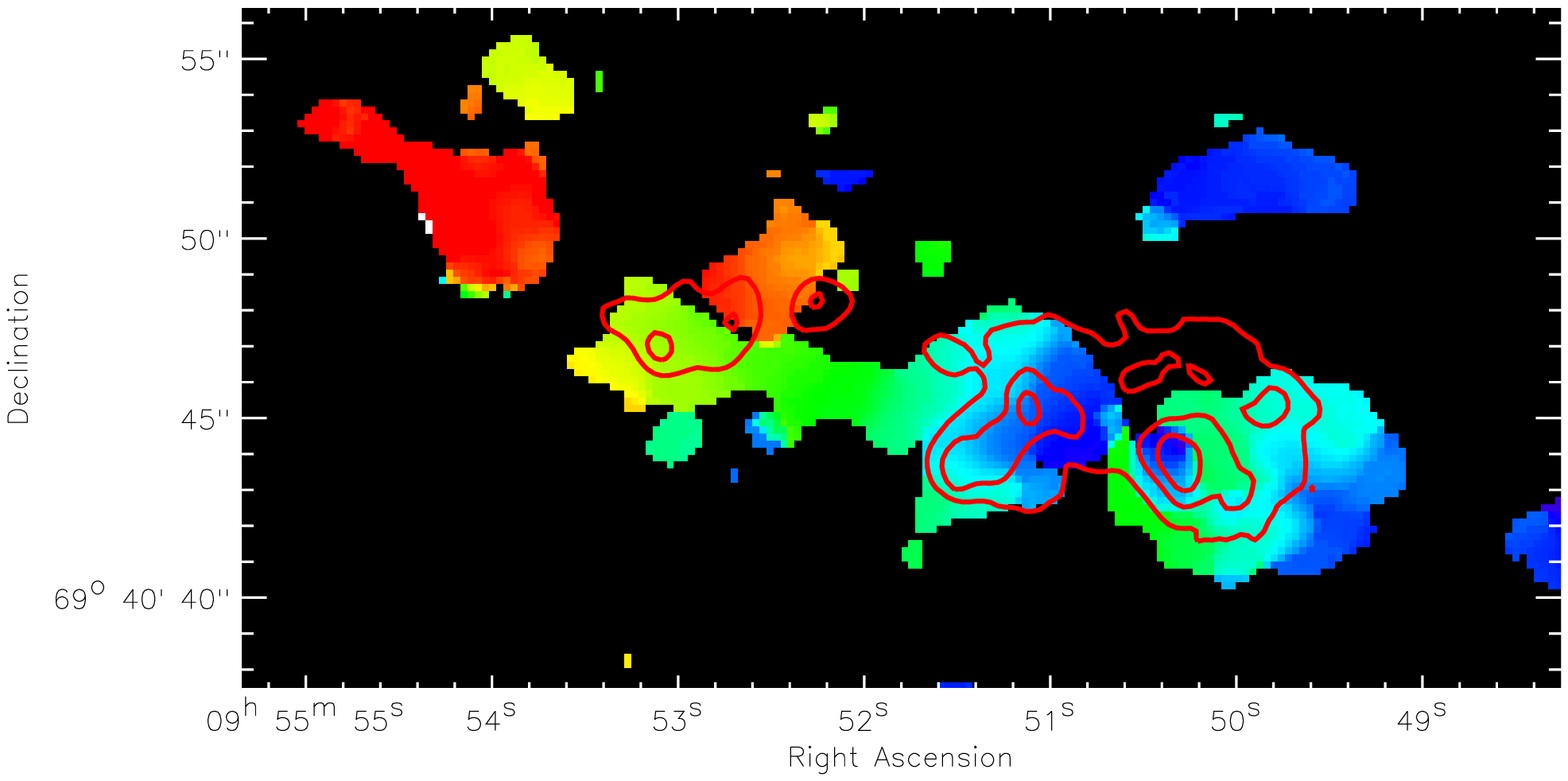}
\vskip -1.0truein
\caption{
\hfill\break
The mid-IR from \cite{LipscyPlavchan2004} in color with contours
of integrated CO(2-1) emission. The contour levels are as in
figure \ref{fig:integrated_intensity}.
\hfill\break
Comparison between CO(2-1) velocity in color and mid-IR emission
in contour. 
The Mid-IR in contour is the same data as the mid-IR in color in the figure above.
The epoch of the coordinates is J2000.
}
\label{fig:midIRvelocity}
\end{figure}

\begin{figure}[t]
\includegraphics[width=4.5truein]{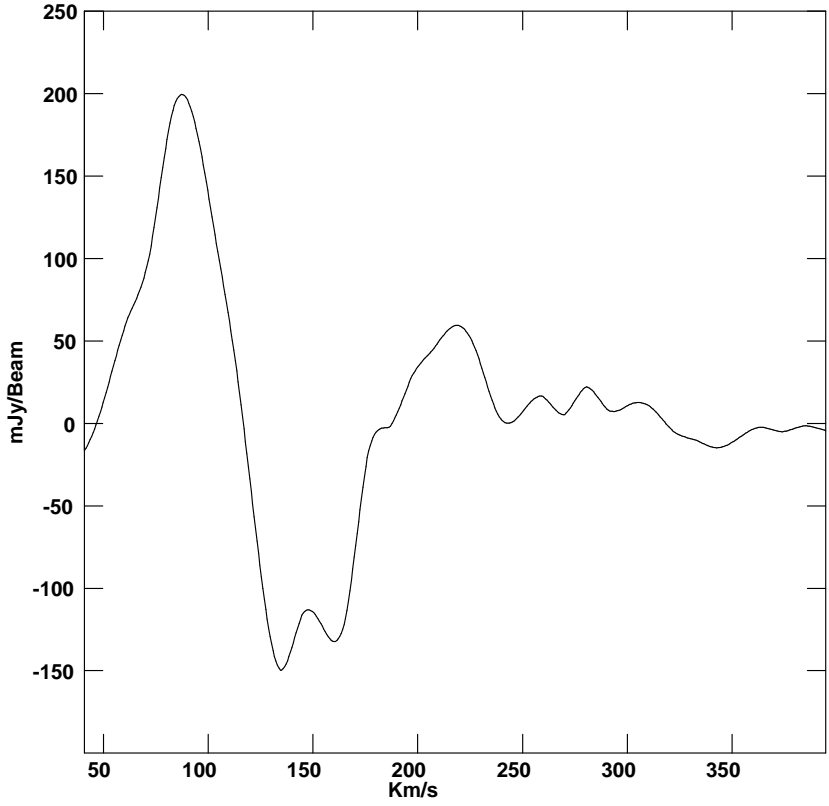}
\caption{
CO spectrum at the mid-IR peak in the small cloud. The inverse P-Cygni spectral line
profile (absorption red-shifted with respect to the emission) 
indicates that the gas is moving inward at a speed of about 35 kms$^{-1}$.
}
\label{fig:spectrumIRpeak}
\end{figure}

The mid-IR and 100 GHz radio continuum identify two regions with the most active
star formation, the large cloud
at 9 h 55 m 51.4 s +69 d 40 ' 44 " (marked A on figure \ref{fig:integrated_intensity})
and the  small cloud to the west (right) at  9 h 55 m 51.4 s +69 d 40 ' 44 " 
(marked B on figure \ref{fig:integrated_intensity}).
Why is star formation most active in these two clouds and not elsewhere?
The observational evidence is consistent with the hypothesis that 
the star formation is triggered by a compressive
shock driven into the molecular clouds. This is determined from the data as follows.
A plot of the intensity weighted average velocity, $\bar v = <Iv>/<I>$ 
shows the velocity gradients in the clouds by the change in $\bar v$ (figure \ref{fig:midIRvelocity}). 
The velocity gradients are strongest where the mid-IR emission is strongest
and the comparison is precise in detail. The smaller cloud with the 
brightest mid-IR emission has a circular pattern of $\bar v$
consistent with radial flow centered on the 
location of the 
strongest mid-IR emission. A spectrum (figure \ref{fig:spectrumIRpeak}) shows
an inverse P-Cygni spectral line profile with red-shifted absorption
and blue-shifted emission. This profile arises from the mixture of ionized gas 
(continuum emission figure \ref{fig:comparison}) and the molecular gas within the cloud.
The red-shifted absorption places the gas moving away from the observer
unambiguously on the near side of the cloud. Therefore this gas is moving
toward the center of the cloud. Similarly the blue-shifted
gas seen in emission must be from behind the ionized gas, on the far side
of the cloud moving toward the observer and toward the center of the cloud. 
The inward velocity is about 35 kms$^{-1}$
as measured by the half the difference in the line center velocities of
emission and absorption.  The velocity field thus shows that this molecular
cloud is radially contracting at highly supersonic velocities consistent with its
being crushed by a sudden increase in external pressure. The spherical pattern
of the flow is consistent with an isotropic external pressure as would be
provided by hot ionized gas.

\begin{figure}[t]
\includegraphics[width=4.5truein,angle=270]{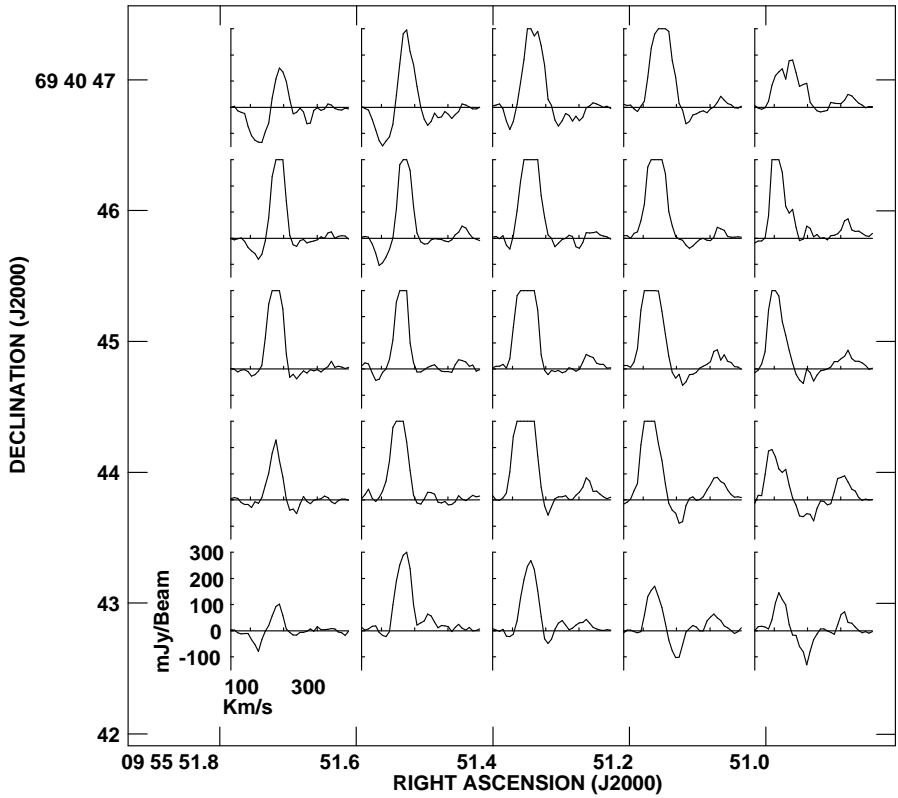}
\caption{
CO spectra across the large cloud. The inverse P-Cygni profiles
(absorption red-shifted with respect to the emission)  seen on the west (right) side of 
the cloud  indicate compression in this region. The classic P-Cygni profiles 
(absorption blue-shifted with respect to the emission) seen on the east (left) side
of the cloud indicate expansion in this east half of the cloud.
}
\label{fig:spectrumMap}
\end{figure}

A comparison of the velocity field in the larger cloud 
(marked A on figure \ref{fig:integrated_intensity})
to the east (left) with
the mid-IR emission again shows that where the gas is in supersonic compression,
the star formation activity (mid-IR emission) is the highest. The ridge of mid-IR emission
is aligned with the lateral velocity gradient seen in the change in $\bar v$ in
figure \ref{fig:midIRvelocity}. Individual spectra in this cloud shows why star formation is
active on the west (right) side of the velocity gradient and the cloud and not on the left.
Figure \ref{fig:spectrumMap} is a map of spectra across the cloud. The spectra from the west (right) 
side of the cloud show inverse P-Cygni profiles with red-shifted absorption 
indicating compression, while the spectra from the 
east (left) side show classic P-Cygni profiles with blue-shifted absorption
indicating expansion.
The inward velocities are similar to those in the smaller cloud, again 20 to 30 km$^{-1}$.  
Thus the velocity field in this cloud shows that where the cloud is in compression, 
star formation is active, and where the cloud is expanding star formation is quiet.
The column density of the molecular gas, proportional to the integrated intensity of
CO, is similar on both sides of the cloud. Thus the compression due to the velocity
field and not the density alone is the difference that is related to the
difference in star formation activity.

The third region of weaker star formation activity 
further east (left) at 9 h 55 m 53 s
+69 d 40m 47 s (marked C on figure \ref{fig:integrated_intensity}), 
shown by the mid-IR emission on figure \ref{fig:midIRvelocity}.
also occurs where the clouds have an internal velocity gradient.
Comparison with the CO integrated intensity figure \ref{fig:integrated_intensity}, indicates that 
the very strong change in $\bar v$ at this position is from the overlap of
two different clouds. The star formation is associated with the velocity gradient 
within each cloud. This gradient is apparent
in the much more subtle change in color across each individual cloud.

Why the most active star formation  is occurring in two particular clouds and not
the other similar clouds may or may not be a significant question. If star 
formation is triggered and sequential, then these clouds may be the current
location
of the progression of star formation.  The other clouds may be
in turn be triggered into compression in the near future. The alternative
is that the location is in some way special in relation to the galaxy, 
apart from the current starburst.

The observation of the velocities of the molecular clouds provides unambiguous
evidence for the causal relationship of the compression of giant molecular
clouds and star formation. 
There is no ambiguity
as to whether the star formation is related to the compression or whether
star formation is proceeding inside the clouds of its own course for example because
of high density. 
In the case of M82, only those clouds that are in compression show active star
formation despite the similarities in density of the largest clouds. In one cloud, 
star formation is occurring in the side under compression
and not in the side that is expanding.  The two sides of the clouds are otherwise
observationally indistinguishable.  Thus the observations imply that star formation
in starbursts is essentially a triggered phenomenon.

\section{Theoretical Considerations and Hypotheses}

The observations suggest that star formation in starbursts is triggered by
the collapse of giant molecular clouds 
driven by a sudden increase in external pressure.  The ionized
gas seen around the molecular clouds is  one possible source of this
pressure as is radiation pressure from a previous generation of stars.  
The observations can be used to understand the origin of this
driving pressure and its more detailed effects on the giant molecular
clouds
by comparing
estimates of the physical conditions in M82 against the conditions
predicted by
different hypotheses about starbursts and cluster formation.  
The results 
of this comparison of theory against observation allow us to describe
a hypothetical general scenario for the starburst process
in galaxies similar to M82.

\subsection{The driving pressure for the collapse of giant molecular clouds}

The ionization may have been caused by strong shocks resulting from cloud collisions
or the pressure of supernovae explosions in a previous generation of starbursts.
A scenario for the ionization of clouds by collisions and the subsequent compression of
giant molecular has been 
outlined by Jog \& Solomon (1992) (JS). They supposed that the collision of
HI clouds in interacting galaxies would lead to their ionization.
Because of their smaller filling factor, the giant molecular clouds would
not collide directly, and as a result they would not be ionized by 
collision, but rather would be
surrounded by hot ionized gas at a pressure substantially higher than
provided by the HI clouds in their previous atomic state. 
Alternatively, Matsushita et al.~(2005)
attribute the ionization of the gas to shocks from a previous starburst episode
that they identify with observed 2.2 $\mu$m emission.
We will examine these two hypotheses with respect to the several observations
of M82.

In their analysis of cloud collisions JS supposed a relative velocity
of 300 kms$^{-1}$,
a typical relative velocity of two galaxies.
The galaxy M82 is not obviously in a direct collision but shows
evidence of gravitational interaction with the neighboring 
galaxy M81 (Yun et al.~1993).
The interaction may still result in the collision of clouds within M82 by the 
distortion of the  rotational equilibrium  that existed
in M82 prior to its close encounter with M81. 
Since the rotational velocity in a typical galactic disk is about
200 km$^{-1}$, and since
our CO observations show a velocity gradient
of about 200 kms$^{-1}$ across the 15" (261 pc) in our map. 
we might expect a relative velocity of this order,
similar to the relative velocity chosen by JS.

In the following analysis, we will assume that shocks
initially heat the gas  to a high temperature and that the post-shock gas is subsequently
rapidly cooled by radiation. 
In discussing the energetics of shocked gas, the simple adiabatic and isothermal 
approximations are useful even though we do not expect the shock to conform
to either of these extremes.
In an adiabatic shock a substantial portion of 
the kinetic energy of the colliding clouds is converted into 
thermal energy of gas. 
In an isothermal shock, all this thermal energy is assumed to be immediately
radiated away so that the gas returns to its pre-shock temperature but at
a higher density. 
We will use the adiabatic approximation to describe
the post-shock gas in its brief hot phase and the isothermal approximation to describe
the post-shock gas after it has cooled. 

We estimate the temperature and density of 
the post-shock gas
before it has had time to cool with the adiabatic approximation
(Spitzer 1978; equations 10-22 and 10-23). In the 
limit of high Mach number, the compression of an adiabatic shock will be 
\be
\rho_2/\rho_1 = \gamma +1 / \gamma - 1
\e{adiabatic_compression}
\ee 
where $\rho$ is the  density and the subscripts 1 and 2 indicate the pre and post-shock gas,
and $\gamma$ is the adiabatic index.
The increase in energy density or pressure, $P_2$, of the gas following the
passage of a shock with velocity, $u_1$ is,
\be
P_2 = 2\rho_1 u_1^2 / \gamma + 1
\e{adiabatic_energy}
\ee
For an initial density in HI of 20 cm$^{-3}$ (JS) and shock velocity equal to the 
relative velocity of 200 kms$^{-1}$, the post shock density and temperature will
be 80 cm$^{-3}$ and $10^6$ K for a gas with an adiabatic index of $\gamma = 5/3$. 
This predicted temperature is the same as the $10^6$ K determined observationally 
from X-ray observations (Griffiths et al.~2000).
The density of the ionized gas may be
estimated from the 100 GHz continuum observations of
Matsushita et al.~(2005). Following Mezger and Henderson (1967),
\be
(n_e/cm^3)^2 = 8\times 10^7 (S_\nu/mJy/beam)(T_e/10^4)^{0.35}(D/kpc)^{-1}
(\nu/GHz)^{0.1}(\theta_x \theta_y \theta_z/arcsec^3)^{-1}
\e{emission_measure}
\ee
where $n_e$ is the electron density, $S_\nu$, is the flux density, $T_e$ is the
electron temperature, $D$ is the distance, and $\theta$ the dimensions of the
emitting region.
The 100 GHz flux  away from the positions of peak emission is 
about 5 mJy/beam (Matusushita et al.~2005).
If the path length is 100 pc, the approximate width of the ionized and X-ray emitting gas, then
$\theta_z \sim 6$".  The width in the map plane $\theta_x \times \theta_y \sim 5\times 4$", twice
the FWHM.
Then the number density would be 90 cm$^{-3}$
comparable 
to the theoretical estimate. 

The shocked gas will cool rapidly, 
\be
t_{cool} = (3/2)nkT/\Lambda
\e{cooling_time}
\ee
where the numerator is the energy density of the gas and $\Lambda$ is the
cooling rate (McKee \& Cowie 1977).
\be
\Lambda = 6.2\times10^{-19} T^{-0.6} n^2
\e{cooling_rate}
\ee
The cooling timescale of the shocked gas is about $10^3$ years.
Thus in the absence of a continuous source of energy to maintain the high
temperature,  one would expect to gas to cool rapidly and we should
observe HI clouds rather than ionized clouds.
If we approximate the properties of the cooled post-shock gas by the isothermal approximation,
the post-shock HI clouds should have a
density equal to the initial density times the Mach number squared.
If the pre-shock HI had an effective sound speed
equal to a typical velocity dispersion of 
1 kms$^{-1}$ (Spitzer 1978), then the Mach number would be 200, and the post-shocked
HI gas would be compressed to an enormously high value. Thus although a single episode of cloud collisions 
would provide conditions similar to those observed, unless we
are observing M82 at the precise moment of the collision, it remains
to be understood how the energy
input from collisions could be maintained.

Let us examine the hypothesis that the ionizing energy is provided by a previous
starburst. Matsushita et al.~(2000) identify a 2.2 $\mu$m peak,
which is located at a position between the two regions of most active,
star formation as emission from an embedded massive star cluster. They estimate
a mass in stars of $2\times 10^6$ M$_\odot$ and 4000 supernovae
explosions over the past Myr for a rate of about 4 supernovae
explosions in a cooling time of 1000 yr. The kinetic energy
release per supernova (type I and II) is $4\times 10^{50}$ ergs  (Spitzer 1978, pg 231;
Blair \& Kirshner 1985)
If this energy goes into a volume of 100 pc$^3$, then the energy
density supplied in a cooling time is $4\times 10^7$ K cm$^{-3}$.
This is close to the estimated energy density ($10^8$ K cm$^{-3}$) 
required to maintain 
the temperature of the ionized gas at $10^6$ K. Thus the energy
input from a single star cluster is sufficient to maintain the
ionized gas at high temperature.

The surface density of
SSCs in areas around the molecular clouds is about
20 per $100^2$ pc$^2$. Thus in the 
area of the ionized gas, there could be more than one cluster, but hidden
behind the extinction of the molecular gas.  Thus the energy available from
supernovae explosions in several clusters is more than sufficient to
maintain the ionization of the ISM.

\subsection{The effect of the pressure on the giant molecular clouds}

The excess pressure of this hot ionized gas will drive a shock into the neighboring
cold molecular clouds at a speed set by the
momentum jump condition,
\be
v_s = (\rho_{ion}/\rho_{mol})^{1/2}c_s
\e{shock_speed}
\ee
Rounding off the estimates in the previous section we may assume that the
energy density of the ionized gas $10^8$ K cm$^{-3}$, the temperature is
$10^6$ K and the density 100 cm$^{-3}$.
The sound speed in the $10^6$ K ionized gas is 90 kms$^{-1}$. 
With an initial
density of 100 cm$^{-3}$ for the uncompressed molecular gas (JS), the shock speed in
the molecular gas will be 60 kms$^{-1}$. The post-shock gas will be accelerated to
some significant fraction of this shock speed, and the observed inward velocities
of 35 kms$^{-1}$ are thus consistent with the estimated shock speed.

In the brief adiabatic phase, the shock will heat  the
gas to a temperature of $2\times 10^5$ with a density of 400 cm$^{-3}$ 
(equations \ref{eq:adiabatic_compression} and \ref{eq:adiabatic_energy})
assuming a molecular weight of 2.33.
The gas will cool rapidly,
molecules will reform, and the increase in density due to the 
shock will be,
\be
n_{final}/n_{GMC} = M^2
\e{isothermal_compression}
\ee
where $M=15$ is the Mach number of the shock assuming $v_s=60$ kms$^{-1}$
and the effective sound speed in the molecular gas is equal to the
velocity dispersion of 4 kms$^{-1}$, appropriate according to the size
linewidth relationship (Larson 1981), for a typical molecular cloud
of 25 pc. 
For the assumed initial molecular density of 100 cm$^{-3}$, the post-shock
density will be $2\times 10^4$ cm$^{-3}$, within a factor of a few the same as  
the average
molecular density of the clouds in our sample $\bar n = 4500 $ 
cm$^{-3}$ 
as determined from
the CO observations.
Because the 
distribution of cloud sizes is a scale free power law, 
the averages reflect only the properties of the clouds
covered by the range of our sample.

The first stars should appear in the post-shock layer in about
a free-fall time,
\be
t_{ff} = ((32/3\pi)G\rho)^{-1/2}
\e{free_fall}
\ee
or $2\times 10^5$ yrs for a density of $2\times 10^4$ cm$^{-3}$. At this time
the layer will be 15 pc thick if the shock speed is 60 kms$^{-1}$.
For our typical molecular cloud with a radius of 25 pc and number density of 100 cm$^{-3}$,
the shocked layer will be in the form of a dense shell at the periphery of the cloud. The mass of the shell will be $3 \times 10^5$ M$_\odot$. 
The star formation rate will
be 
\be
SFR = M_{stars}/t_{ff}
\e{SFR}
\ee
If we assume a
star formation efficiency of 50\%, the minimum required to form a bound star
cluster, the star formation rate in the 
compressed shell of one cloud will
be  0.6 M$_\odot$ yr$^{-1}$.  
This is also approximately the star formation rate for the
whole cloud if the shock progresses to the center of the cloud.

\subsection{Star formation by radiative compression}

Suppose the pressure that is driving the shock is reduced if the hot ionized gas
surrounding the molecular cloud cools or 
if its density is reduced by outflow in a galactic wind.  
Star formation may continue nonetheless because
the radiation pressure from the first stars
that are formed in the post-shock gas  will be sufficient to continue
the compression. If the first star formation in an interacting galaxy
is initiated by the collision and ionization of HI clouds as suggested by JS, 
then the radiation pressure from the first stars formed may allow star formation
to continue through a cloud 
despite the short time scale for the collisional ionization discussed
above and in JS.

In the example in the section above, the total luminosity  of the stars in the compressed layer 
will be (JS; Scoville \& Young 1983)
\be
L_{tot} = 1.3\times 10^{10} (L_\odot yr/M_\odot) SFR
\e{Ltotal}
\ee
or $8\times 10^9$ L$_\odot$. This luminosity will generate
an inward radiation pressure of
$ 4.5 \times 10^8$ K cm$^{-3}$, assuming the pressure is given by,
\be
 P = (1/2) L/(4\pi r^2 c)
\e{radiation_pressure}
\ee
with $c$ the speed of light.
Thus the radiation pressure
from the first wave of star formation is of the order  of the 
pressure of the ionized gas that began the compression of the
molecular cloud. Therefore the star formation
will now be inwardly self-propagating.
The inward speed of the wave of star formation
will be equivalent to the initial shock speed owing to the similarity
in the pressure of the ionized gas and the radiation pressure of the shell
of star formation. Thus once star formation is triggered by compression,
the star formation will move inward at supersonic velocities even if
the external pressure is rapidly reduced.  The possibility of self-propagating
star formation triggered by radiation pressure 
is a possible means around the problem discussed
in JS of the short time scale for compression due to cloud collisions.

\subsection{Fragmentation by  thermal and gravitational instabilities}

If a high Mach number shock is driven into a molecular cloud so that
the cloud is temporarily ionized, then upon cooling, the gas may
fragment by thermal and gravitational instabilities.  Because the
temperature enters into the equation for the cooling rate (equation \ref{eq:cooling_rate})
as a negative power, the gas is thermally unstable under both isobaric
and isochoric conditions (equation 4, Field (1965)).
Fragmentation by the thermal instability is especially interesting because
it allows for a fragmentation on time and length scales that would shorter than
those expected from gravitational
collapse.

The thermal instability will be most effective in fragmenting the
gas if the timescale of the thermal instability is comparable to the
dynamical time scale of a fragment. 
The growth rate of the thermal instability is roughly equal to the cooling rate
of the gas. If the post-shock gas is heated to $2\times 10^5$ K and
compressed to 400 cm$^{-3}$, then the cooling time (equation \ref{eq:cooling_time})
is only 10 yr. The dynamical time scale is given by a length divided by the sound speed.
The length scale of a density perturbation
that would give rise to a condensation of
a solar mass  would be about 0.3 pc, as given by,
\be
\lambda = (3M/4\pi \rho)^{1/3} 
\e{length_scale}
\ee
Assuming a sound speed of 100 kms$^{-1}$ appropriate for gas at $2\times 10^5$ K, 
the dynamical time scale for a perturbation of this size $\lambda/c_s \sim 3000$ yrs.
The disparity in time scales implies that
the gas will cool under isochoric conditions. Thus the thermal instability as 
described by linear perturbation analysis will be 
unable to compress the gas in the colder regions before all the gas has cooled and
the thermal instability ceases.  However, non-linear effects may allow the thermal
instability to play a role in fragmentation (Murray \& Lin 1989; Murray \& Lin 1991;
Inutsuka \& Koyama 2004; Baek et al.~2005). 
Fragmentation by non-linear effects
must be investigated by numerical methods beyond the scope of this paper.

Although the gravitational or Jeans instability in purely spherical geometry is
unable to fragment a  cloud, if there are pre-existing density perturbations in the
cloud, the passage of the shock and subsequent compression of the gas may make some
of these perturbations unstable to gravitational collapse.
Once the post-shock gas has cooled, it will be significantly 
denser, about $2\times 10^4$ cm$^{-3}$, than in the pre-shock state, 100 cm$^{-3}$. 
At this density, the size of a solar mass perturbation will
be $\lambda \sim 0.07 ({\rm M/M}_\odot)^{-1/3}$ pc. The escape speed at the boundary of the perturbation
will be, $ v_{\rm esc} = (GM/\lambda)^{1/2}$ or 0.25 (M/M$_\odot$)$^{1/3}$ kms$^{-1}$. The perturbation
will be unstable by the Jeans criterion if this escape speed exceeds the sound speed or
effective sound speed of the molecular gas. What affect the shock will have on the velocity dispersion of
the molecular gas is not clear. 
If the
internal velocity dispersion of the post-shock gas is the same, 4 kms$^{-1}$, as in the
pre-shock state, then
a condensation of a few 100 M$_\odot$ will be unstable, 
and the compression of the cloud
reduces the Jeans mass to at least 100 times smaller than in the pre-shock cloud
where the Jeans mass was approximately the size of the entire pre-shock cloud. 

The assumption that the sound speed of the cloud remains the same in the compressed
cloud as in the original cloud results in a higher Jeans mass than in typical clouds because
the compressed cloud has an internal velocity dispersion of the much larger pre-shock
cloud and the
compressed cloud then no longer follows the standard size-line width relation
(Larson 1981). 
If the effective sound speed is reduced on passage through the shock then
the Jeans mass will
be lower than 100 M$_\odot$. 

The process of fragmentation has interesting implications for properties of stars
within clusters. For example, if fragmentation were to take place very rapidly
as might be achieved if fragmentation were accelerated by the thermal instability,
then the stars within the cluster might more likely have the same metallicity because
they would be formed at the same time from the same molecular gas.  If fragmentation
takes place gravitationally, then if the Jeans
mass of the compressed cloud were higher than in typical molecular clouds, the
initial mass function (IMF) might be different in an SSC than in a typical open
cluster in our galaxy.

\subsection{The dominant mode of star formation in starbursts}

Are all stars in a starburst formed in clusters? The star formation
rate proceeds as the 1.5 power of the surface density according to the
Schmidt law or according to simple scaling arguments, as the volume 
density divided by the free-fall time, $\rho/t_{ff}$.
If the shock compression of the molecular gas is a factor of about 100,
then the star formation rate in a shock compressed cloud
will be 1000 times faster than in a normal giant molecular cloud.
Thus most of the stars formed in a starburst will form in clusters.

Because the cloud mass spectrum is a power law extending at least
to clouds of a few thousand M$_\odot$, we would expect the
compression of many small clouds and their transformation into clusters. 
These small clusters
with their relatively weak gravitational binding energy might not 
survive very long against disruptive tidal forces in the crowded 
star burst environment. The dispersal of these cluster would
liberate stars to the field. However with a power law mass spectrum
most of the mass is in the largest clouds, and most of the stars
in the starburst would remain in massive clusters.

\subsection{A universal process for the formation of open and globular clusters}

A fundamental question about the SSCs is whether
the formation of these gravitationally bound massive clusters is as suggested by the
hypothesis of shock compressed giant molecular clouds
fundamentally different than the
processes that we see operating in our own Galaxy, or whether
a bound massive star cluster could be formed simply by a scaled up
version of normal star formation.  
In cluster formation in our Galaxy, with a few possible exceptions
(e.g.~Clark et al.~2005),  
the formation of the first few massive OB stars tends to disperse
the gas of the host giant molecular cloud resulting in a low efficiency
for star formation and leaving insufficient stellar mass
for the cluster to remain gravitationally bound.
However,  in the higher density environment of
a starburst, because the star formation rate scales with the gas
density as the $\sim 1.5$ power according to the Schmidt law 
(Schmidt 1959; Kennicutt 1998), perhaps the rate of star formation might be
high enough to transform most of the cloud into stars before
the dispersal of the molecular gas.
Elmegreen \& Efremov (1997) (EE) have described just such a model
for cluster formation with the additional hypothesis that
the high pressure environment will also result in a lower
dispersal rate that  improves the efficiency. The lower rate of dispersal
results from the higher gravitational binding energy of the
clouds that in turn derives from
the assumption of virial equilibrium in a high pressure environment.
A cloud in equilibrium in a high pressure environment must have 
a higher internal energy $\rho c_s^2$.
To maintain virial equilibrium, the cloud must also
have a higher gravitational binding energy
that should result
in a lower rate of dispersal.

In this model, the gas in a cloud is diminished by two processes: transformation to
stars and dispersal by the energy of the newly formed
stars. Star formation proceeds until there is
no more gas in  the cloud. The efficiency of star formation
then depends on the relative rates of star formation
and dispersal. Following EE, if the rate of star formation is proportional
to some power of the density and the rate of dispersal
is proportional to the stellar luminosity and inversely
proportional to the gravitational binding energy of the
cloud, then high density clouds will evidently form
stars more efficiently. By scaling the rates of star formation
and efficiency from those in the solar neigborhood,
EE are able to show that a cloud in an environment with
a pressure 1000 times higher, a cloud of $10^4$ M$_\odot$
is able to form stars at an efficiency of 50\%. This
efficiency of course depends on the cloud beginning its
star formation and dispersal at its assumed density and
pressure. In other words, in a high density, high pressure
environment, either the cloud is formed at that density and
pressure, or star formation must be delayed until the cloud
reaches a threshold density and pressure. 

\begin{figure}[t]
\includegraphics[width=4.5truein]{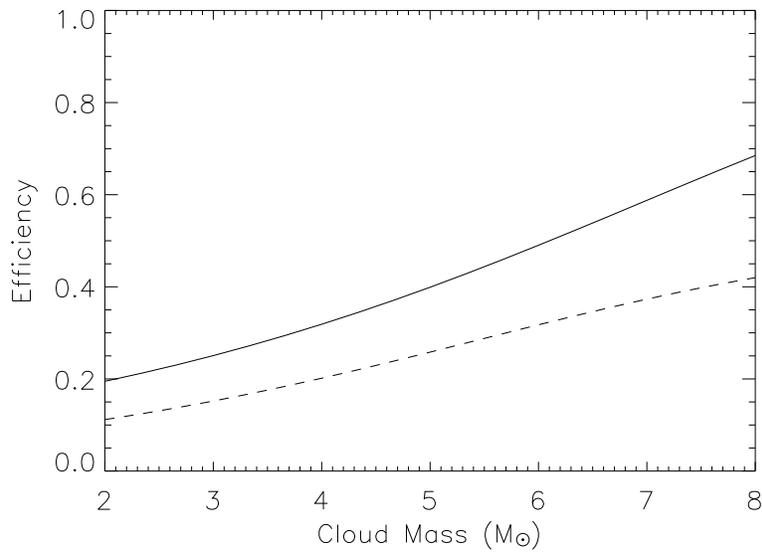}
\caption{
The solid line is the calculation exactly from Elmegreen and Efremov (1997) that
shows that the efficiency of star formation as a function of the mass of clouds 
in an environment with
a pressure 160 times higher than that in the solar neighborhood. The dashed line
is an improved calculation that shows that starting from normal pressure
and increasing the pressure over a crossing time lowers the efficiency.
High efficiency requires forming the cloud at high pressure  or a threshold
to prevent star formation at lower pressures.
}
\label{fig:EE}
\end{figure}

However, if the cloud is compressed from a lower density, then
the star formation efficiency will be lower.
For example, the rate of change of the mass of the gas in
a cluster is given in EE by their equation (3),
\be
dM_c / dt = dM_s/dt - A L/c_t^2
\e{cluster_static}
\ee
where $M_c$ is the mass of gas, $dM_s/dt$ is the star formation
rate, and the last term is the rate of dispersal proportional
to the total luminosity of stars in the cluster and inversely proportional to the
internal turbulent velocity dispersion, $c_t$.
If however, a low density cloud that is allowed to form stars,
is brought to a higher pressure, always assuming virial equilibrium,
then the equation for the rate of change of the mass of the
gas in the cloud will be,
\be
dM_c / dt = dM_s/dt(1- \alpha t)(1 - A L/c_t^2)(1 - \alpha t)^{-1/4}
\e{cluster_dynamic}
\ee
where $\alpha$ is the rate of increase of the pressure,
the star formation rate scales directly with the pressure, and
the dispersal rate scales as $P^{-1/4}$. The latter two scalings are
derived as in EE from virial equilibrium and the Schmidt law, 
$dM_s/dt \propto \rho^{1.5} \propto (P^{3/4})^{1.4} \sim P$ and
$c_t^2 \propto M^{1/2}P^{1/4}$. The difference in the result of 
the two equations is plotted in figure \ref{fig:EE}. The top
line shows the efficiency for clouds of different masses 
that start forming stars 
at a pressure of 160 times
that in the solar neighborhood. The lower dashed line shows the
efficiencies for clouds that are brought 
up to this pressure over a crossing time and are allowed to
form stars along the way as the pressure increases. 
Since these latter clouds
spend some time at lower pressures, even the most massive clouds never
reach an efficiency of 50\% and will therefore never form 
a massive gravitationally bound cluster. 

The pressure increase of 160 was chosen to match the conditions in M82.
The ratio of the average density of the M82 molecular clouds to
those typical in our Galaxy is about 45. 
Since in this theory the density scales as the 3/4 power of the pressure,
the observed  molecular density in our M82 clouds would correspond to an increase in
pressure by a factor of about 160. At this overpressure, clouds
of $10^5$ M$_\odot$ would have an efficiency of about 40\% 
for the static case and 20\% for the case of slowly increasing pressures.
Thus star clusters produced in clouds of mass less than $10^5$ M$_\odot$ 
would  be less likely to be gravitationally bound,  and most would begin to disperse within
a crossing time. Our sample of clouds shows a power law mass
spectrum from $10^6$ down to below $10^4$ M$_\odot$. For a sample of
star clusters with ages greater than a crossing time, this
theory would predict a power law mass spectrum of slope $-1.5$ for star clusters
with masses greater than the minimum mass required for gravitational binding,
about $10^6$ M$_\odot$ and a declining 
cluster mass spectrum for masses below this value. Thus the 
cluster mass spectrum would  be peaked at $10^6$ M$_\odot$ for the
case of a static overpressure, and few bound clusters would be expected
for the case
of increasing pressure.

We can estimate the dispersal time of the stars in a cluster from its 
velocity dispersion and radius
of a cluster. Observations of a few clusters indicate velocities of 10 to 15 kms$^{-1}$
for clusters with radii of a 1 and 2 pc and masses of $10^5$ M$_\odot$ (Ho and Filippenko 1996)
The implied crossing and dispersal times  are then about $10^5$ years.
The stellar clusters in the nearby region of M82 known as B1 and B2
(deGrijs et al.~2000) have ages greater than  $10^8$ yr and show equal
numbers of SSCs detected above and below $10^5$ M$_\odot$. 
If the clusters are unbound below a mass of $10^6$ M$_\odot$, we would
expect very few unbound clusters to survive for the age of the B1 and B2 
regions which is more than 1000 crossing times.
If the observed mass distribution were corrected for completeness 
(Whitmore et al.~1999), the number of low mass clusters in M82 B1 and B2
would be higher implying
the formation of many bound clusters below the mass limit
predicted by the theory of universal cluster formation. 
Thus the observations are
inconsistent with the hypothesis of cluster 
formation by conventional (Galactic) star formation
in a high pressure environment.

This result bears on the hypothesis of super-giant
molecular clouds \citep{HarrisPudritz1994, Wilson2003}. If the hypothesis
of universal cluster formation in a high pressure environment is not satisfactory,
then there is no motivation for supposing the existence of super-giant 
molecular clouds to
provide the high pressure around internal giant-molecular-cloud-sized clumps
that will transform into massive star clusters. 
Furthermore, we do not detect super-giant molecular clouds in our observations.
The largest cloud detected in our sample is less than $10^7$ M$_\odot$.
Our observations 
are sensitive to structure as large as about 170 pc. 
A cloud of
this dimension with density, 4500 cm$^{-3}$ equal to the average density
of clouds in our sample, would have a mass of $10^8$ M$_\odot$. Thus we should
be able to detect clouds up to $10^8$ M$_\odot$, but perhaps not larger.
Thus there is no motivation or evidence for super-giant molecular clouds in M82. 

The hypothesis that individual SSC's are formed from individual giant molecular clouds
is consistent with the observational finding that SSC's are found in complexes or
clusters of SSC's \citep{Whitmore1999, Melo2005}. The implication then is that the individual
molecular clouds were found in complexes prior to compression. This implied structure of the ISM is  
consistent with a number of theoretical conceptions about the ISM, for
example the idea that the clouds are structures in a hierarchical turbulent cascade,
or that the structure of the ISM has similarities with a fractal structure.  Both
turbulent and fractal structures are scale-free, but would appear clustered when
not completely resolved by observations. Thus 
a complex of normal giant molecular clouds that would be transformed into a complex of SSC's
could appear as a super giant molecular cloud when viewed
with limited angular resolution. 

The difference between the hypothesis of cluster formation within super giant molecular
clouds and the hypothesis of cluster formation through shock 
compression as suggested by our observations
is not so much a difference in the scale of the clouds but of the process of compression.
If the clouds are cores within a larger super giant molecular cloud and 
compressed by a surrounding  molecular cloud, then these cloud-cores
would
be compressed by molecular gas. In the alternative
hypothesis of shock compression by the pressure of ionized gas
as proposed for M82, the clouds are compressed by surrounding ionized gas. 
Secondly, although the hypothesis of formation within super giant molecular clouds
does not specifically propose a time scale, if one wishes the process to remain 
consistent with a universal process of cluster formation, then the compression  of a cloud-core
within the super giant molecular cloud should not take places on a time scale so 
much shorter than
the crossing time of the core, that  a high velocity shock 
develops within the core. Star formation by shock compression
is not the way star formation proceeds
at lower pressures in our own Galaxy. Thus in the process of universal
cluster formation, there is an implied minimum time scale of about a crossing time. In contrast,
the time scale for cluster formation by shock driven compression could be much shorter, for example,
the shock crossing time, or the free-fall time of the Jeans mass scale fragments
discussed in the section on fragmentation.

\section{Conclusions}

We have resolved the ISM in the M82 starburst into 300 molecular clouds. The
mass spectrum has a slope of $-1.5$ consistent with the unevolved upper end
of the mass spectrum of
super star clusters in M82 and of globular clusters other galaxies.
The internal velocities of the giant molecular clouds with the most active 
star formation indicate massive compression on the scale of the entire cloud.
Simple theoretical considerations suggest that  the star formation
is triggered by the compression of a shock driven into the molecular
clouds by the pressure of the surrounding ionized gas. 
The wave of star formation may
be self-propagating because the radiation pressure from the first stars
formed is sufficient to continue the compression of the molecular cloud.
This mode of star formation appears quite different from normal star
formation in our own Galaxy, and in particular star formation in
starbursts does not appear to be a scaled up version of star formation
as we see in open clusters in our own Galaxy.

\acknowledgements
The authors thank George Field and Doug Lin for their help in exploring
the interesting possibility  of the thermal instability.
L. C. H. acknowledges financial support from the Carnegie Institution of 
Washington and by NASA grants from the Space Telescope Science Institute 
(operated by AURA, Inc., under NASA contract NAS5-26555).

\clearpage

\end{document}